\documentclass[12pt,preprint]{aastex}
\newcommand{\8}{\ensuremath{^{48}}Ti}

\slugcomment{}

\shorttitle{Isotopic Titanium Abundances}
\shortauthors{Chavez \& Lambert}

\begin{document}


\title{Isotopic Titanium Abundances in Local M Dwarfs} 


\author{Joy Chavez}
\affil{Department of Astronomy, University of Texas,
       1 University Station, C1400, Austin, TX 78712-0259}
\email{jchavez@astro.as.utexas.edu}
\and
\author{David. L. Lambert}
\affil{McDonald Observatory, University of Texas at Austin,
       1 University Station, C1402,  Austin, TX, 78712-0259}
\email{dll@astro.as.utexas.edu}


\begin{abstract}

Relative abundances of the five stable isotopes  of titanium
($^{46}$Ti to $^{50}$Ti)
are measured for 11 M dwarfs belonging to the thin disk (four stars),
thick disk (three stars), the halo (one star), and  either
the thick or the thin disk (three stars). Over the metallicity
range of the sample ($-1<$[Fe/H]$<0$), the isotopic ratios are approximately
constant to the solar system ratios. There is no discernible difference
between the isotopic ratios for thin and thick disk stars. Isotopic
ratios are in fair
accord with recent calculations of Galactic chemical evolution despite the
fact that such  calculations underpredict [Ti/Fe] by about 0.4 dex
at all metallicities. 

\end{abstract}


\section{Introduction}

Insights into the chemical evolution of the Galaxy are provided by the
relative abundances of elements in stars belonging to
 the main populations of the
Galaxy: halo, disk (thick and thin), and bulge. In a common representation
of abundances, one plots [X/Fe] versus [Fe/H] for the element X and
Fe.\footnote{Standard notation is used here: [A/B] = $\log$(A/B)$_{\rm star} -
\log$(A/B)$_{\odot}$ and $\log A\equiv \log\epsilon(A) = \log$(N$_{\rm A}/$N$_{\rm H})
+ 12.00$ where $N$ is the number density.}
  Elements
may be grouped into classes with each element in a
class exhibiting a similar behavior.  Comprising one such class are the
$\alpha$-elements:  the elements in this
class include Mg, Si, S and Ca, each with its most
abundant isotope having a mass number that is  a multiple of four.
Observations show that the runs of [X/Fe] for each of the $\alpha$-elements are
similar for stars of a given population presently in the solar neighborhood.
The value of [X/Fe] is positive and
constant (say, $+0.3$ dex) for [Fe/H] $< -1$ but decreases 
to the solar value (i.e., [X/Fe] $=0$) as thin disk stars are sampled at
[Fe/H] $> -1$. The change in the variation of [X/Fe] versus  [Fe/H]
for $\alpha$-elements  at [Fe/H] $\simeq -1$ is widely attributed to the
onset of contamination of interstellar gas with the Fe-rich $\alpha$-poor
ejecta from Type Ia supernovae that reduces the [$\alpha$/Fe] previously
established by $\alpha$-rich Fe-poor ejecta from Type II supernovae.
Theoretical modeling of Galactic chemical evolution (GCE) reproduces quite
well the [$\alpha$/Fe] versus [Fe/H] trends for the standard $\alpha$
elements - see, for example,
 Timmes, Woosley, \& Weaver (1995, hereafter, TWW95),
Goswami \& Prantzos (2000, hereafter, GP00), and
Kobayashi et al. (2006,  hereafter,
K06). Such modeling enterprises,  however, ignore the observational differences
between the runs of [X/Fe] versus [Fe/H] for thin and thick disk stars.

Titanium with its principal isotope being
 $^{48}$Ti would appear to  qualify as an
$\alpha$ element. Indeed, the observed run of [Ti/Fe] with [Fe/H]
is similar to that of a standard $\alpha$-element. Yet,
GCE models fail to predict the observed trend. For example, K06
 who reproduce satisfactorily the observed trends for Mg, Si, and
Ca fail to match the trend for Ti; the predicted Ti trend has the
form expected for an $\alpha$ element  but at all [Fe/H], including the
solar value, the predicted [Ti/Fe] is about 0.4 dex below the observed
trend. GP00 and TWW95 report  comparable
discrepancies: the predicted [Ti/Fe] are 0.3 to 0.5 dex below the
observations for models that predict satisfactorily the run of
[Ca/Fe] with [Fe/H]. In general, the K06 predictions are a better fit to
observations for elements from C to Zn than either the TWW95 or
GP00 predictions. These failures to account for the evolution of the
Ti abundance are not attributable to inappropriate choices for the
reference (i.e., solar) abundances of Ti and Fe in confronting
predictios with observations. Uncertainties over the solar Ti and
Fe abundances cannot erase a 0.4 dex difference in [Ti/Fe] by revisions
of the zero-points for [Ti/Fe] and [Fe/H]. For example, substitution
of Asplund et al.'s (2005) Ti and Fe abundances for those of Anders \&
Grevesse (1989) used by K06 shifts K06's predictions by 0.22 dex to
higher [Fe/H] and by 0.13 dex to higher [Ti/Fe], a far cry short of the
0.4 dex failure.
 Clearly, aspects of Ti nucleosynthesis are not yet understood.

Recent work on the compositions of thick and thin disk stars (e.g.,
Bensby et al. 2005; Reddy, Lambert, \& Allende Prieto 2006)
resolve the previously
considered single [$\alpha$/Fe] versus [Fe/H] relation for disk stars
into different relations for thick and thin disk stars. Titanium behaves
like Mg, Si, and Ca in its [X/Fe] versus [Fe/H] differences for the
thick and thin disks. Modeling of the thick and thin disks has yet
to reach the detailed treatments given in the above cited and other
papers for the chemical evolution of the  halo-disk combination but
apportionment of disk stars between the thin and thick disk is not
going to solve the 0.4 dex  Ti problem.

With the intent of providing novel observational evidence on titanium
nucleosynthesis, we have measured relative abundances for the five stable
Ti isotopes for a selection of M dwarfs drawn from the halo, thick and
thin disks and spanning iron abundances from the solar value to
about [Fe/H] $= -1$. The dominant isotope is $^{48}$Ti, the
prospective $\alpha$-nuclide. The other isotopes are $^{46}$Ti, $^{47}$Ti,
$^{49}$Ti and $^{50}$Ti with solar system abundances of 8.25, 7.44, 5.41,
and 5.19 per cent, respectively, and with $^{48}$Ti accounting for the
lion's share at 73.72 per cent (Lodders 2003). Predictions about the variation
of the isotopic abundances with [Fe/H] depend primarily on the yields
from and the relative frequencies of
Type II and Type Ia supernovae.

Pioneering predictions of the variation of the Ti isotopic ratios  with [Fe/H]
were provided by TWW95: relative to $^{48}$Ti and the solar
system isotopic ratios, the predictions for $^{46}$Ti, $^{47}$Ti, $^{49}$Ti
and $^{50}$Ti were factors of two too large, three too small, spot on, and
two too small, respectively. Isotopic ratios were predicted to
decline  with decreasing [Fe/H]  by factors of eight for
$^{46}$Ti, six for $^{47}$Ti, two for $^{49}$Ti, and 30 for $^{50}$Ti
between [Fe/H]$=0$ and $-1$. These predictions and, in particular, more
recent examples invite observational tests. 
Measurements of the isotopic abundances in stars of different
metallicities may not resolve directly
the 0.4 dex Ti problem but have the potential to suggest directions
in which to look for its solution.

In this paper, we derive the isotopic abundances
from high-resolution spectra of M dwarfs
covering the TiO molecule's $\gamma$-system's 0-0 band with its leading
red-degraded bandhead at 7054 \AA. The chosen spectral window
was previously used in investigations of sunspot spectra by
Lambert \& Mallia (1972), 
 Mira by Wyckoff \& Wehinger (1972),  $\alpha$ Tau by Lambert \& Luck (1977),
 and  a sample of late-type dwarfs and giants by Clegg, Lambert, \& Bell (1979).
These previous exploratory analyses
 provided neither the precision for the isotopic ratios
nor the coverage in [Fe/H]  necessary to
subject predictions  to a quantitative
test. Our results  provide the first quantitative tests
of the predictions for disk stars.

\section{Selection of stars}

Three criteria were applied to the selection of the program stars. First,
in order that TiO
lines in the $\gamma$-system 0-0 band be of sufficient strength to provide
detectable lines of the four less abundant Ti isotopes, dwarf
stars with  a spectral type of
early
M were chosen. Dwarfs were prefered to giants
 because  their sharp lines  allow
clear resolution of lines of the different isotopes; giants provide
broader lines resulting in a merging of the lines of different
isotopes (Clegg et al. 1979). Second, a magnitude limit was necessarily
applied  in order that an adequate signal-to-noise be obtained in a
reasonable total exposure time at the resolving power of 120,000. 
Third, a probability calculation was applied to identify members of the
thin and thick disks and the halo.

The population assignments were made using the probability recipe
suggested by Bensby et al. (2003, 2005) and applied also by
Reddy et al. (2006 - see equations 1 \& 2) The recipe
requires
(i)  the Galactic velocities $U,V,W$
($U$ is the velocity toward the center of the
Galaxy, $V$  the velocity in the direction of Galactic rotation, and $W$  the
velocity toward the north Galactic pole) corrected to the local standard of
 rest, and (ii) the mean $U,V,W$ and their dispersions of
the thin disk, thick disk, and halo in the solar neighborhood, and (iii)
the relative stellar densities of the three populations.
  Reid, Gizis, \& Hawley (2002)
provide a catalog of $U,V,W$  heliocentric velocities for M dwarfs
 which we correct to the
Local Standard of Rest (Dehnen \& Binney 1998).
The descriptions of the
kinematics and relative
densities of the three populations are those adopted by
Ram\'{\i}rez, Allende Prieto, \& Lambert (2007).

%TABLE 1
Table 1 provides the population assignments for our stars.
The  stars include one halo star, three thick disk stars, four thin
disk
stars and three that might belong to either the thin or the thick disk.
To be identified with a particular population, we required
 the membership probability to be greater than 75\%.  Otherwise, we considered
the star to belong to either the thin or the thick disk.

\section{Observations and Data Reduction}
\label{obs}

High-resolution spectra of the TiO $\gamma$-system's 0-0 band
from 7045 \AA\ to 7094 \AA\ were obtained for the  chosen M dwarfs
with the W.J. McDonald Observatory's 2.7m Harlan J. Smith
Telescope and its {\it 2dcoud\'{e}} spectrograph (Tull et al. 1995)
at a resolving power of about 120,000 for all stars except LHS178 for
which a resolving power of 60,000 was used.
Observations were made in four observing runs between July and December 2006.

For the chosen configuration of the cross-dispersed echelle spectrograph,
about 25 \AA\ of an order were recorded for about 20 orders
with central wavelengths from 6060 \AA\ to 9400 \AA. Two settings
were required to cover the 7045 \AA\ to 7094 \AA\ interval: a blue setting
covered 7045 \AA\ to 7073 \AA, and a red setting covered
7067 \AA\ to 7094 \AA.  Exposures of 30 minutes each were coadded
as necessary to realize the desired S/N ratio:  S/N ratios just to the blue of
the 7054 \AA\ bandhead range from 90 for LHS178 to 250 for GJ880. 
 A Th-Ar hollow cathode
lamp was observed to provide the wavelength calibration
 and a measure of the instrumental profile.  A rapidly-rotating
hot star was observed for telluric line removal and correction for the
echelle blaze. Standard IRAF\footnote{IRAF
is distributed by the National Optical Astronomy Observatories, which are
operated by the Association of Universities for Research in Astronomy, Inc.,
 under cooperative agreement with the National Science Foundation}
 reduction techniques were used.
Since the TiO lines of interest are distributed across the blue and
red settings, it was necessary to merge the two spectra to provide
a continuous run from shortward of the leading red-degraded bandhead
of the 0-0 band at 7054 \AA\ to about 7090 \AA.
The region 7045 \AA\ to 7054 \AA\ was used to set the continuum.
An order  in the merged spectrum
 covering  8420 \AA\ to 8470 \AA\ provides a set of
Ti\,{\sc i} lines used in the determination of the Ti abundance.

\section{Method}
\label{Method -- Ingredients}

Isotopic abundance determination involves
fitting synthetic spectra to an observed spectrum. Key ingredients
needed for computation of a synthetic spectrum are a suite of
appropriate model atmospheres, a line list for TiO and other
contributors, a set of stellar parameters (effective temperature,
surface gravity, metallicity, microturbulence ($\xi$),
 macroturbulence ($\zeta$), and projected
rotational velocity),
and a code for the computation of the synthetic spectra. In this latter
context, we use the program MOOG (Sneden 1973) which assumes LTE and
considers consistently ionization and association of atoms into
molecules.
 In subsequent
sections, we describe the employed ingredients.

\subsection{Model Atmospheres}

Model atmospheres were taken from the  NEXTGEN  grid (Hauschildt, Allard, \&
Baron 1999)
 used by  Bean et al. (2006) with an interpolation in
effective temperature ($T_{\rm eff}$), surface gravity ($\log$ g),
and metallicity ([Fe/H]). The NEXTGEN models assume a
microturbulence ($\xi$) of 2 km s$^{-1}$. 
The adopted composition for the models is a
scaled solar composition (i.e., [X/Fe] = 0) and the solar Fe abundance
is taken as $\log\epsilon$(Fe) = 7.45.

\subsection{Line List}

The TiO lines are drawn from Plez's (1998) who provides data for
the leading electronic transitions: wavelengths, excitation potentials,
and $gf$-values. His list includes lines for all five Ti isotopes in combination
with $^{16}$O.
For our primary region (7045 \AA\ to 7085 \AA),  we imposed a cut-off
in strength in order to reject lines that
make a negligible contribution to the stellar spectrum. The final list of
more than 5500 lines contains in addition to the 0-0 band P, Q, and R
branch lines for the five isotopic varieties of TiO, satellite
branch lines from the 0-0 band of $^{48}$TiO, and $^{48}$TiO lines of
$\Delta v$ = +1 $\gamma$-system bands.\footnote{Lines from Ti$^{17}$O
 and Ti$^{18}$O are not included in the linelist on account of the 
anticipated low abundances of these isotopes; their abundances are
likely less than those of the solar system for which $^{16}$O/$^{17}$O
= 2700 and $^{16}$O/$^{18}$O = 480.
The complex of stellar TiO lines
will include very  weak contributions, probably
 irretrievably blended, from $^{48}$Ti$^{18}$O for the coolest dwarfs.}
 It will be shown below
that, as contributors to the stellar spectra,
 the satellite branches and the hot bands  rank
 with
the lines of the
0-0 main branch lines from the four less abundant isotopes.
 A laboratory spectrum of TiO described by Davis, Phillips, \&
Littleton  (1986) was
retrieved from the NSO library.\footnote{\url{http://solarch.tuc.noao.edu/diglib/query\_by.html}}
This 
spectrum was
used to check the wavelengths of the main branch 0-0 lines for all
five isotopic varieties. 
%Atomic lines were taken from
%Kurucz \citep{Kurlines}.

The Plez line list includes many more lines
than we have included for the computation of synthetic spectra. 
For inclusion in our line list, lines had to pass a simple test.
The strength of each line was estimated from the following
relation:\[S=\log(agf\lambda)-\theta\chi,\] where $a$ is abundance,
$gf$ is the $gf$-value, $\chi$
is the excitation energy, and $\theta=5040/T_{eff}$. For
this estimate, a typical but cool M dwarf temperature of 3300 K was
applied.  To test for the significant contributers to the TiO spectral
region, several syntheses were performed, starting with an extremely
stringent cutoff, allowing only the strongest lines.  This cutoff was
reduced by 0.5 dex until the change in the spectra was less than 0.5\%.
Weaker lines  were left out of the syntheses.

The Plez list was also used to provide the TiO lines, primarily from
the $\epsilon$-system's 0-0 band that contaminate the 8420 \AA\ to
8470 \AA\ interval that provides a  determination of
 the Ti abundance from Ti\,{\sc i}
lines; the 0-0 band's red-degraded head is at 8446 \AA.
For the Ti\,{\sc i} lines, we computed solar $gf$-values using the
solar flux spectrum (Kurucz et al.  1984), the Kurucz (1993) model solar
atmosphere, the microturbulence $\xi = 1.13$ km s$^{-1}$
(Grupp 2004), and a solar Ti abundance of $\log\epsilon$(Ti) = 4.90
(Asplund, Grevesse, \& Sauval 2005).

In all cases, we adopt without alteration, Plez's choices for the
$gf$-values based on a combination of laboratory measurements,
primarily radiative lifetimes, and quantum chemistry calculations.
We also adopt Plez's choice for the TiO
dissociation energy of TiO: D$_0 =  6.87^{+0.07}_{-0.05}$
eV, the value determined by experiment (Naulin, Hedgecock, \& Costes 1997).

\subsection{Stellar parameters}

Our principal goal -- the determination of the relative
abundances of the five Ti isotopes -- is, as we show below, insensitive
to the adopted effective temperature, gravity and metallicity
of the adopted model atmosphere. A key parameter is the
microturbulence because the abundance ratio of a trace
isotope to $^{48}$Ti is 10 to 20 to 1 or greater and, thus, the
$^{48}$TiO line may be saturated when its isotopic counterparts are
conveniently detectable.  However, the other stellar parameters are important
in determining the metallicity and it is through the variation
of the isotopic abundances with metallicity that we test
the predictions of Galactic chemical evolution. Adopted values for the
stellar parameters are given in Table 2 with the discussion in the following
subsections providing the  adopted methods.
%TABLE 2%

\subsubsection{Effective temperature}

Initial estimates of $T_{\rm eff}$ are based on a star's spectral
type  (Reid, Hawley, \& Gizis 1995) and the  spectral type - effective
temperature relation provided by Reid \& Hawley (2005). This procedure
is anticipated to provide $T_{\rm eff}$ to about 100 K.
Small adjustments to $T_{\rm eff}$ were made in some cases in order that
the Ti abundance from the Ti\,{\sc i} and TiO lines were consistent.
For Gl 699, we adopt $T_{\rm eff}$ from Dawson \& DeRobertis (2004)
based upon estimates of the star's angular diameter and total flux.

\subsubsection{Surface gravity}

The M$_K$-mass relationship from Delfosse  et al. (2000) was exploited as
a route to the surface gravity
because it has a weak metallicity dependence. This relation with
Bean et al.'s (2006) $\log\ g$ - mass relation provides an
estimate of the surface gravity. {\it Hipparcos} parallaxes (Perryman
et al. 1997)
 and 2MASS $K$ magnitudes (Cutri et al. 2003) are combined
to give M$_K$. Only LHS 178 is not in the {\it Hipparcos} catalog
and for this star we take the parallax from Gliese \& Jahreiss  (1991).
Typically, the $\log\ g$ is estimated to about $\pm$0.01 dex but this
does not include the systematic errors resulting from the use of the
two relationships.

\subsubsection{Microturbulence}

The ideal approach would be to determine $\xi$ from the TiO
spectrum. Across the observed stretch of the $\gamma$-system's 0-0
band, the $^{48}$TiO lines from the P, Q, and R lines do not
show a sufficient difference in strength to provide a
useful estimate of $\xi$. Additionally, many lines are blended with
lines of the other isotopic varieties. In principle, one may use the
satellite $^{48}$TiO lines for the desired comparison of weak
and strong lines from which to determine $\xi$. However, the
spectrum is such that examples of clean satellite lines
are impossible to find; prospective satellite lines are
blended with isotopic lines of similar strength from the
P, Q, and R lines, and also P, Q, and R lines of
the  $^{48}$TiO $\gamma$-system's hot ($\Delta v= +1$) bands.

As an alternative to comparing strengths of weak and strong
lines of the same molecular or atomic species, we bound
$\xi$ from the widths of TiO lines. An upper limit to $\xi$
is determinable from imposing syntheses for different $\xi$ on an
observed spectrum. After synthetic spectra are convolved with the
instrumental profile appropriate for that observation,  
the upper limits with a slight star-to-star variation run from
1.5 km s$^{-1}$ to 2.5 km s$^{-1}$. These are
upper limits because we neglected contributions to the line width
from rotation and macroturbulence.

Our assumption regarding macroturbulence is that it equals
microturbulence, an empirical result of approximate validity
for the Sun and main
sequence stars but taken as an extrapolation  for late-K and early-M dwarfs.
Rotational velocities of M dwarfs are generally very low
(Reiners 2007), say $v\sin i < 1$ km s$^{-1}$ equivalent to
about a 0.2 km s$^{-1}$ apparent contribution to $\xi$.
With our assumptions - macroturbulence equals microturbulence and no
rotation - the microturbulence for each star is provided.
 These values of $\xi$ are
a factor of $\sqrt{2}$ smaller than the upper limits. The
uncertainty is about $\pm0.5$ km s$^{-1}$. Saturation of lines
is controlled, of course, by the quadratic combination of the
thermal velocity and the microturbulence.  The thermal velocity of a
TiO molecule at a representative temperature, say 3000 K, is
$\sqrt{2kT/m} = 0.9$ km s$^{-1}$,
and, thus, a microturbulence less than about 1 km s$^{-1}$ has
but a small effect on the saturation of lines.
In some cases, small adjustments were made to the microturbulence
on the basis of fits to the profiles of the Ti\,{\sc i} and TiO  lines.
The microturbulence is not reliably derivable from the small suite of
Ti\,{\sc i} lines used to obtain the Ti abundance; there are  no
weak Ti\,{\sc i} lines (the molecular haze impedes the measurement of weak
lines), and the measured lines cover too limited a range in
strength. 

Our results for the microturbulence and macroturbulence
 are completely in line with results from the recent
spectroscopic analyses of Woolf \& Wallerstein (2005) and
Bean et al. (2006). For the early M dwarfs, here T$_{\rm eff} < 4000$ K,
Woolf \& Wallerstein obtained a mean value of $\xi = 1.0$ km s$^{-1}$ from
19 stars; their estimates are from Ti\,{\sc i} lines and
the usual constraint that the Ti abundance be independent
of equivalent width. Bean et al. determined microturbulence ($\xi$) and
macroturbulence ($\eta$) from a fitting procedure to a suite of
atomic line profiles. Their results from five stars gave mean values:
$\xi$ = 0.9 km s$^{-1}$ and $\eta$ = 1.0 km s$^{-1}$; note that our
assumption $\xi = \eta$ is essentially verified by these
results. These $\xi$ determinations from atomic lines are consistent
with our measurements reported in Table 2.

\subsubsection{Metallicity}

An   estimate of the metallicity is obtained
from the Ti\,{\sc i} lines,  their
solar $gf$-values and the microturbulence values just
discussed  with an
iteration such that the input metallicity for the model atmosphere
is equal to that derived from the lines.  Four Ti\,{\sc i} lines are used in the
final determination: 8457.103 \AA\ from multiplet 174 with $\log gf = -1.85$,
8476.147 \AA\ from multiplet 182 with $\log gf = -1.26$, and 8438.923 \AA\ and
8450.892 \AA\ from multiplet  223 with $\log gf = -0.79$ and $-0.84$,
respectively. Multiplet numbers are from Moore (1945).
The lines in  order by strongest to weakest are 8439, 8451, 8457, and 8467 \AA.
Other Ti\,{\sc i} lines are in our spectral window but were rejected either
because they are seriously blended or are stronger than the above quartet and
then sensitive to the (uncertain) damping constants.
 The [Fe/H]  is estimated
from the [Ti/H] using mean trends for the [Ti/Fe] versus [Fe/H] for thin and
thick disk stars (Reddy et al. 2006): [Ti/Fe] = $-0.18$[Fe/H] for the
thin disk, and [Ti/Fe] = $-0.03$[Fe/H] $+0.2$ for thick disk and halo
stars.

In determining the Ti abundance from the TiO lines, we constrain the
analysis by adopting relationships between the C, O, and Fe abundances.
The [Ti/Fe] relation was as above. The [C/Fe] and [O/Fe] relations,
also from Reddy et al. (2006), are: [C/Fe] $= -0.23$[Fe/H] for the thin and
thick disk and halo stars, and [O/Fe] $= -0.19$[Fe/H] for  the
thin disk and [O/Fe] $= -0.25$[Fe/H] for thick disk and halo stars.
Adopted solar abundance are $\log\epsilon$(C)=8.39 and
$\log\epsilon$(O)=8.66 (Asplund et al. 2005).

These assumptions about the composition are not entirely consistent
with  the adoption of a scaled solar composition
used for the computation of the model atmospheres. This inconsistency
will have a negligible effect on the derived Ti isotopic abundances.

%\include{method}                     
%\include{Results}

%\section{Method -- results}
\label{sec:Results}
\section{The metallicity of the M dwarfs}

The Ti\,{\sc i} lines were each fit using the fitting procedure
applied by Bean et al. (2006) assuming the microturbulence and
macroturbulence velocities discussed above. 
Results for the four lines are
%TABLE 3
given in Table 3.
% with profile fits for
%selected stars shown in Figures \ref{ap16Gl699}, \ref{ap16Gl701}, and 
%\ref{ap16LHS178}.
This analysis assumes Local Thermodynamic Equilibrium (LTE).
Results from the four lines are generally consistent: 8467 \AA, the
weakest line, gives an abundance that is about 0.1 dex less than the
average, and the 8457 \AA\ line gives an abundance about 0.1 dex more than
the average. These small differences have a negligible effect on an
interpretation of the trend of isotopic abundances with metallicity.
Perhaps, larger effects in the line-to-line scatter and the mean
Ti abundance result from our assumption of LTE - see Hauschildt et al. (1997)
for a discussion of non-LTE effects on the Ti\,{\sc i}
spectrum in M dwarfs.

The $^{48}$TiO 0-0 band lines were fit to obtain a Ti abundance with
the results given in Table 3. The abundance there
tabulated includes a correction for the four trace isotopes in order
that a direct comparison may be made with the abundance from the atomic
lines which necessarily refers to the total Ti
abundance because the isotopic wavelength shifts are negligible.

Agreement between the Ti abundances obtained from atomic and molecular
lines is good except for LHS 178. In part, this is a result
of small  adjustments to $T_{\rm eff}$ (see above) for a few stars: adjustments
were 100 K or less except for LHS 2018 where $T_{\rm eff}$ was
lowered by 250 K from the value indicated by the spectral type.
The [Ti/H] are converted to [Fe/H] by the recipe given above.

The high-resolution spectroscopic model atmosphere (NEXTGEN)
 analysis of Fe\,{\sc i} and Ti\,{\sc i}
lines for  GJ701 by Woolf \& Wallerstein (2005) gave
[Fe/H] $= -0.20$ and [Ti/H] $= -0.25$  in good agreement with our
results (Table 3). For LHS178, our  Ti\,{\sc i} lines give [Ti/H] $=-0.97$
and  TiO lines [Ti/H]$=-0.55$ where the disagreement in part reflects a low
quality of fit to the atomic lines. We adopt the average  value which
translates to [Fe/H] $=-1.0$.  
The only other estimate of the metallicity of LHS 178 is from  photometric
 band strength indices of (Gizis 1997).  These indices are a measure of
CaH and TiO band heads.
The degeneracy of temperature and abundance can be broken with
 the double-metal, temperature sensitive TiO against the single-metal, less
 temperature sensitive molecule of CaH.  
This  technique
 gave [Fe/H]$= -1.0\pm$0.5, which   agrees with our
 measurement.

Our
determinations of the iron abundance [Fe/H]
are supported by the photometric calibration of the
metallicities of M dwarfs by Bonfils et al. (2005).
Their calibration of [Fe/H] in terms of a polynomial expansion involving the
absolute magnitude $M_K$ and the color index $(V-K)$ was provided by
combining two datasets. Spectroscopic [Fe/H] abundances obtained by
Woolf \& Wallerstein (2005) for late-K and early-M dwarfs provide about half of
the calibrators. The other half are dwarfs with spectral types from
K7 to M6   belonging to wide visual binaries in which the primary is
a F, G, or K dwarf whose spectrum is amenable to an abundance analysis
by standard spectroscopic techniques. The cool secondary is assumed - reasonably
so - to have the [Fe/H] of its primary companion. These 48 calibrators
provide the polynomial expansion.\footnote{Johnson \& Apps (2009)
remark that the photometric calibration underestimated [Fe/H] (relative
to spectroscopic estimates) for metal-rich M dwarfs. This effect is
unimportant here; our smaple does not include metal-rich ([Fe/H] $>$ 0)
stars.}
%TABLE 4
Table 4  and Figure  1 show the comparison between our results and the
values given by the Bonfils et al.'s calibration. The agreement
is satisfactory. A possibly discrepant point is that for GJ 699 but
for this star the $M_K$ is outside the limits of the calibration.
%Other metallicity estimates for individual stars are discussed
%in Section \ref{Stars}.

A photometric technique for the determination of $T_{\rm eff}$ and metallicity
has been developed by Casagrande, Flynn, \& Bessell (2008)
 using Johnson-Cousins and
2MASS near-infrared photometry. Luca Casagrande (private communication)
has kindly applied their technique to our stellar sample.  
A comparison with our results is given in Table 4 and Figure 2.
The effective temperatures obtained by Casagrande et al. agree within 100 K
with ours except for LHS2018 and LHS1226, a result generally consistent
with the expected errors of the two techniques.  Interpreting the
photometric metallicity [M/H] as [Fe/H], these results agree quite well
with ours. Casagrande et al. note agreement between their [M/H] and
recent determinations at the 0.2 dex level in general. Eight of our 11 stars
match the photometric metallicity to within $\pm0.2$ dex.  The exceptions
show differences (Us - Casagrande) of $-0.3, +0.4$ and $-0.7$ dex where only the
latter for LHS2018 might be considered a concern. For each of these
three, Bonfils et al.'s recipe gives results in  good
agreement with ours.

In summary, the [Fe/H] in Table 3 appear reliable at the
$\pm0.2$ dex level. These values define well the x-axis in the
plots of isotopic ratios versus metallicity that are used to test
predictions of GCE.

\section{The isotopic abundances}

Isotopic abundance ratios are derived by fitting synthetic spectra to an
observed spectrum. Ratios of the four lesser abundant isotopes
with respect to $^{48}$Ti are sensitive primarily to the microturbulence.
Ratios among the four lesser abundant isotopes are insensitive to the
microturbulence. Both forms for expressing the isotopic ratios are
sensitive to blends chiefly  from  weak $^{48}$TiO lines including the
satellite transitions from the same 0-0 band that provides the
main lines of interest and from hot bands of the $\gamma$-system. In the
following subsections, we discuss the determinations of isotopic
ratios.

\subsection{Observed and synthetic spectra - General remarks}

The strength of the TiO lines changes appreciably, as expected,
across the stellar sample. These lines are strongest in GJ 699 and
weakest in GJ 215. Lines from the four trace isotopes are prominent in
stars with strong TiO yet clearly present in sample stars with
weak TiO. Isotopic abundances are determined from matching
synthetic to observed spectra.
%FIGURE 4
%FIGURE 5
%FIGURE 6
Figures 3 to 5 show these spectra for representative
stars: GJ 699 with strong TiO, GJ 701 with medium strength TiO,
and LHS 178 with weaker TiO (and our lowest metallicity star).

In Figure 3 for GJ 699, the abundance of \8 is fixed and synthetic
spectra for  three different
isotopic fractions $^{i}$Ti/\8 are shown. By inspection, one
can identify the features to which one or more of the trace isotopes
contributes. The measurement of isotopic fractions is discussed
below. 
Inspection also suffices to show wavelength regions for which
the synthetic spectra fail to match the observed spectrum. In those
cases  where the observed spectrum  shows the greater absorption, one presumes
that unidentified lines not included in the line list used for
computing the synthetic spectrum  are depressing the spectrum. These
regions are ignored in assessing the isotopic
fractions. Surprisingly, there are places where the synthetic
spectra shows more absorption than the observed spectrum.
%These discrepancies are discussed below.

The TiO molecule's contribution to the observed spectra is not
simply from the $\gamma$-system's 0-0 band and its P, Q, and R lines
and their five isotopic components. The 0-0 band satellite branches
contribute lines. In addition, the $\gamma$-system's $\Delta v = +1$ sequence
contributes lines. Our inclusion of satellite and $\Delta v = +1$ lines
adds `noise' to the spectrum. By way of
%FIGURE 5
illustration, we show in Figure 6 for GJ699 the separate contributions to the
synthetic spectrum of (i) the TiO 0-0 band P, Q, and R lines from all
isotopes, (ii) the 0-0 band satellite \8O lines, and (iii) \8O
lines from the $\Delta v = +1$ bands. It is obvious that contributions
(ii) and (iii) are comparable to that from the four trace isotopes in (i).
Strengths and positions of the satellite lines are most probably reliably
represented in Plez's line list. Some of the $\Delta v = +1$ lines have
not been recorded on laboratory spectra and their predicted wavelengths
use molecular constants beyond the range in which they have been
established. In addition, their $gf$-values are subject to greater
uncertainty than are the values for the 0-0 band. It is because of these two
latter uncertainties that the $\Delta v = +1$ lines contribute
unwanted noise. 
Our list contains only about six atomic lines and these we do not identify
separately.

%A similar dissection of the TiO contributions is shown in Figure \ref{Gl701hotsat}
%for GJ 701, a star with weaker TiO and a hotter effective temperature.
%Again the satellite and $\Delta v = +1$ lines provide noise with the latter
%the more important contributor at this effective temperature.
%{\bf Joy A GJ 701 figure like that for GJ 699} {\it Done.}

No synthetic spectrum is ever a perfect fit to an observed spectrum.
Our synthetic spectra are no exception. 
Failures of the synthetic spectra may be put in two classes: (i) the
observed spectrum is stronger than the synthetic spectrum, and (ii) and
the reverse of this where the synthetic spectrum shows absorption stronger
than in observed. In principle, both (i) ands (ii) are open to
simple and obvious interpretations. As noted above, (i) admits of the
possibility that the adopted line list is missing lines or the strengths of
 included lines are underestimated (i.e., adopted $gf$-values  are too
small). Similarly, (ii) may have a simple explanation, i.e., the adopted
$gf$-values are too large.  

The impossibility or indeed the inevitability of composing a thoroughly
complete  line list means that one must accept in any comparison 
occurrences of class (i) and (ii) failures. Given the very high
quality fit of synthetic to observed spectra, as demonstrated in Figures 3, 4,
and 5, a low frequency of failures cannot surely adversely affect determinations
of the isotopic ratios. 
Our impression is that wavelength errors for the lines from the
$\Delta v = +1$ bands may be largely responsible for the failures of
class (i) and (ii).
 There are, however, places where the synthetic
spectrum with just the 0-0 main lines (all five isotopes) is stronger
than the observed spectrum. Two such examples are seen in Figure 6 at
7074.45 \AA\ and 7079.65 \AA. Scrutiny of the line list and the laboratory
spectrum of TiO shows that the TiO lines from the less abundant isotopes
are at their measured wavelengths and there are no atomic lines in
the line list that are contributing unwanted absorption. Furthermore, these
discrepancies are seen across the sample at the same (stellar)
wavelengths and are not, therefore,
attributable to noise, incorrect correction for telluric (H$_2$O)
absorption lines or emission (OH) from the night sky. 
One may wonder if these discrepancies arise from {\it stellar}
emission lines. 

\subsection{The Isotopic Fractions}

In the final fitting of synthetic spectra to an observed spectrum,
the microturbulence and macroturbulence are held fixed. A determination
of the \8 abundance from a fit to the least-blended \8O
features is made. Below, we discuss the uncertainty in this
abundance arising -- principally -- from the influence of the microturbulence
on the saturation of these lines. Next, the isotopic fractions are
determined.

An automated fitting routine was adapted for determining these
fractions. For the final determinations, an inspection of the fit between
the observed spectrum and
synthetic spectra for several different  isotopic mixtures
was made of selected features. 
%(Table \ref{isofeatures}).
%TABLE 5
The selection (Table 5) emphasises those features
to which one or just two isotopes contribute. For the 
lesser abundant isotopes,
we started with the \7O dominated features of which there were
several. For each an abundance for $^{47}$TiO
 was determined and a weight assigned
from the
quality of the best fit.
The weighted average abundance of \7O and its dispersion was calculated
and then
adopted for the analysis of blends to which \7O is a contributor. Next, we
considered \6O features in the same manner. Finally, \9O and
\0O were considered. Isotope ratios are simply obtained from these
molecular abundances.  An inspection of the complete wavelength region was
conducted to check for the quality of the fit with the synthetic spectrum
computed for the final isotopic abundances.
%TABLE 6

Table 6 illustrates this procedure for
the three representative stars GJ 699, GJ 701, and LHS 178.
The feature-to-feature scatter in the abundance of a given trace isotope
reflects partly the noise in the observed spectrum and partly the
complexity of the spectrum and in particular contributions from
satellite and $\Delta v = +1$ lines.
Table 7 summarizes the Ti isotopic abundances for each of the
program stars. In Table 8, the isotopic abundances are expressed as
a fraction of the total (all isotopes) Ti.
%TABLE 7
% The isotopic fractions relative to
%\8 and the relative abundances of the four trace isotopes are given
%in Table \ref{fractions}{\it??  But these are relative to the total abundance.  Do you want me to change this to ratios wrt to \8?}.
  The uncertainties in Table 7 and 8 calculated from the line-to-line
scatter are one contributor to the total uncertainty; systematic
errors are discussed in the next section.

LHS178 obviously enjoys a special place in the interpretation of the GCE
predictions; it is the most metal-poor star and the sole
representative of the Galactic halo in our small sample. It was, as noted
above, the only star observed at a resolving power of 60,000 and not
120,000. Figure 5 shows three synthetic spectra and the
observed spectrum. The spectrum for pure $^{48}$TiO including
satellite lines and lines from the $\Delta v=+1$ bands does not fit the
observed spectrum, and, therefore, it is difficult to escape the
conclusion that the less abundant Ti isotopes have left their
imprint on the observed spectrum. A comparison of observed spectra for
other stars with TiO strengths similar to those for LHS178 suggests
too that the less abundant isotopes are contributors to the LHS178 spectrum.
Nonetheless, their abundances are plotted as upper limits in Figures 7 and
8.

\subsection{Systematic errors}
\label{sec:Errors}

 The relative abundances of trace
isotopes are only very weakly dependent on the adopted stellar
parameters. This is not the case for the ratios with respect to \8
because the \8O lines are saturated, quite severely so for
stars like GJ 699 and less severely so for GJ 701; the ratio of the
depths of the \8O to \7O lines is smaller than the approximately
12 to 1 ratio of the abundances. Saturation brings into prominence the
sensitivity of the measured isotopic fractions (abundances relative to
\8) to the microturbulence.

Errors introduced to the isotopic ratios by errors in the atmospheric
parameters were characterized through a series of syntheses starting with
a synthetic  spectrum for the standard atmospheric parameters
 with 1\% noise added.  Three different stars were
represented: a cool
star much like GJ699, a medium temperature star like GJ701, and a warm
star mimicking the TiO-weak stars.  With these as the ``observed'' spectra,
the chi-squared minimization routine found the best fit isotope abundance
as each of the parameters was changed. 
Adjustments to T$_{\rm eff}$ of $\pm100$K and to $\log$ g of $\pm0.5$ dex
resulted in nearly constant changes to the abundances of each of the
varieties $^{i}$TiO with the result that the isotopic abundance changes
were less than about 1 to 2\%, changes less than the errors arising from the
line-to-line scatter. The abundance 
changes were approximately $\pm$0.2 dex for the
temperature change and $\pm$0.09 dex for the gravity change. An [Fe/H]
change of $\pm$0.2 dex in the adopted value for NEXTGEN model
atmosphere introduces a small change in isotopic abundances and 
negligible changes in the isotopic fractions.

 Adjustments to the adopted microturbulence $\xi$ from an assumed
value of 1 km s$^{-1}$ and across the uncertainty range of
$\pm0.5$ km s$^{-1}$ provide small changes in the isotopic
fractions when referenced to the $^{48}$Ti or total Ti
abundance for all stars except those with the strongest TiO lines
(e.g., GJ699). For GJ701, for example, the $^{48}$TiO abundance was
decreased by about 0.03 dex for the increase in $\xi$ from 1.0 to
1.5 km s$^{-1}$ with isotopic abundance ratios relative to $^{48}$Ti
changing by 0.02 dex or less. The $\xi$ increase has a greater effect for
the GJ699-like stars with the $^{48}$TiO abundance decreasing by 0.10
dex and the isotopic ratios relative to $^{48}$Ti increasing by
about 0.06 dex. These changes are of opposite sign and smaller for 
the reduction in $\xi$ by 0.5 km s$^{-1}$ because of the contribution
of the thermal velocities to the total velocity controlling the saturation
of the lines. These systematic uncertainties arising from $\xi$ are no larger,
even smaller than, the random errors from the fitting of the synthetic
spectra and the blending arising from the satellite and $\Delta v =+1$
lines. Certainly, the ratios among the four lesser abundant isotopes
are dominated by random and not systematic errors.

\section{Discussion}
\label{Discussion}

\subsection{Observed trends}

Our results are summarized in Figures 7 and 8. In
Figure 7, we display the isotopic fractions $^{i}$Ti/$^{48}$Ti
as a function of [Fe/H] with the different symbols denoting thin (open
squares), thick (filled circles) disk stars and the cross the halo star
LHS 178. The three stars with kinematics that do not allow a
clean attribution to either the thin or thick disk are
represented by half-filled circles. In Figure 8, we display ratios
among the less abundant isotopes with $^{46}$Ti in the
denominator, and also the ratio $^{46}$Ti/$^{48}$Ti.
% $^{46}$Ti  seems least sensitive to
%the details of the nucleosynthesis accomplished by supernovae, at least
%redictions by GP00 and K06 are in good agreement for this
%isotope in contrast to the poor agreement for the other three
%less abundant isotopes. 

Three statements suffice to summarize the results in  Figure 7.
First, the stars with [Fe/H] $\sim 0$ display the
solar system isotopic ratios, as expected by every other
abundance measure for the Sun and stars of solar
metallicity;  mismatches between the GCE predictions and the solar
system abundances are not attributable to the latter being anomalous.
 Second, there is no clear difference between
isotopic abundances for thin and thick disk stars across this small
sample.  Third, the isotopic ratios are
sensibly constant over the observed metallicity range from [Fe/H] of
zero to about $-0.8$.
A similar set of  conclusions applies to Figure 8 where ratios
with respect to $^{46}$Ti are given.

\subsection{Nucleosynthesis predictions}

Pioneering predictions of the variation of the Ti isotopic ratios  with [Fe/H]
were provided by TWW95: relative to $^{48}$Ti and the solar
system isotopic ratios, the predictions for $^{46}$Ti, $^{47}$Ti, $^{49}$Ti
and $^{50}$Ti were factors of two too large, three too small, spot on, and
two too small, respectively. Isotopic ratios were predicted to
decline steeply with decreasing [Fe/H]: declines by factors of eight for
$^{46}$Ti, six for $^{47}$Ti, two for $^{49}$Ti, and 30 for $^{50}$Ti
between [Fe/H]$=0$ and $-1$. 

 Such predictions do not match our results
terribly well and, in particular, the [Fe/H] dependences appear at odds
with the observations. More recent predictions provide a closer
fit to the observations and given the complexity of
calculations of nucleosynthetic yields by Type II and Ia supernovae and
certain ingredients in a GCE recipe (i.e., the initial
mass function or  IMF), factors of
two agreement between prediction and observation should be considered
a success. Two recent predictions are shown in Figure 7: 
(i) GCE as represented by  K06 with data supplied by
Kobayashi (2008, private communication), and (ii) GCE as modeled by
GP00 with updates contributed by Prantzos (2008,
private communication).
 Before remarking
on the comparison between these two predictions and our observations, we
discuss the yields of the five Ti isotopes from both types of
supernovae.

Isotopes of titanium are synthesized in both Type II and Type Ia
supernovae. Predictions of the evolution of the isotopic abundances
with metallicity involve calculations of the relative
frequencies of  Type II
and Ia supernovae and their respective yields
 plus a gaggle of additional assumptions about
stars (e.g., the IMF) and the Galaxy (e.g.,
infall rates). Our intent here is not to diagnose critically the
published predictions for the relative isotopic abundances but rather
to extract from published
predictions aspects of the nucleosynthesis of the Ti isotopes.
%Three set of predictions will be called upon in our
%discussion: Timmes et al. (1995), Goswami \& Prantzos (2000), and
%Kobayashi et al. (2006).

K06 (their Table 3) present yields (in solar masses) from
stellar generations of Type II supernovae and hypernovae with
initial metal mass fractions for $Z = 0.0, 0.001, 0.004$ and 0.02.
The predicted yields expressed as  number density ratios are
$^{46}$Ti/$^{48}$Ti = 0.055 (0.093), $^{47}$Ti/$^{48}$Ti = 0.055 (0.050),
$^{49}$Ti/$^{48}$Ti = 0.035 (0.052), and
$^{50}$Ti/$^{48}$Ti = 0.004 (0.053) for initial
compositions $Z = 0.001$ and in parentheses $Z = 0.02$. (The value $Z = 0.02$
is now a suprasolar value: the solar value is $Z = 0.012$ according to
Asplund et al. (2005).)
 Apart from the case of $^{50}$Ti, the yields
relative to $^{48}$Ti are weakly dependent on the initial $Z$ but $^{50}$Ti
relative to $^{48}$Ti is produced below the one per cent level for $Z$ less
 than about 0.003.

Other calculations of Type II supernovae yields predict a steeper
rise in the isotopic ratios with increasing $Z$.
Woosley \& Weaver (1995) and Chieffi \& Limongi (2004)  presented 
yields in ejecta as a function of $Z$ after  integration 
over an IMF (with other assumptions) showing
 a factor of about ten increase
for $^{46}$Ti and $^{47}$Ti, a factor of two increase for
$^{49}$Ti, and a factor of 30 increase  for $^{50}$Ti (all
relative to $^{48}$Ti) from $Z=0.1Z_\odot$ to $Z=Z_\odot$
(Prantzos, private communication). For $^{46}$Ti and $^{47}$Ti, these
increases are much larger than those given by K06. 

Type Ia supernovae also contribute to the Ti abundances.
For the yields from Type Ia supernovae, TWW95  took
the model W7 for all $Z$ with the nucleosynthesis as calculated by Thielemann,
Nomoto, \& Yokoi (1986): W7 is a popular model (Nomoto, Thielemann, \&
Yokoi, 1984).
 This model gives ratios with respect to
$^{48}$Ti of 0.099 ($^{46}$Ti), 0.0016 ($^{47}$Ti), 0.046 ($^{49}$Ti), and
0.0077 ($^{50}$Ti). K06's calculations also adopt the model W7 for all $Z$
but with the ejecta's composition as calculated by Nomoto, Thielemann, \&
Yokoi (1994) and Nomoto et al. (1997): the ejecta  has  
ratios with respect to
$^{48}$Ti of 0.088 ($^{46}$Ti), 0.0030 ($^{47}$Ti), 0.082 $(^{49}$Ti), and
0.060 ($^{50}$Ti). Note the nearly tenfold increase in the $^{50}$Ti
relative abundance over TWW95's adopted value.  

The predictions by Prantzos in Figure 7 are based on the prescription for
GCE described by 
GP00 who took yields for Type II supernovae
from TWW95. For the predictions in Figures 7 and 8, Prantzos
took  yields from Chieffi \&
Limongi (2004) but the changes in the isotopic ratios attributable
to the switch in Type II supernovae yields are slight.
GP00 adopted yields
for Type Ia supernovae
from Iwamoto et al. (1999) for two models: W7
and W70. The former assumes the exploding
white dwarf evolved from a star of solar $Z$ and the latter that the
white dwarf's progenitor had $Z = 0$. For their GCE calculations, GP00
interpolate linearly in $Z$ to obtain yields as a function of $Z$.
The W70 (W7) yields provide distinctly non-solar isotopic fractions:
$^{46}$Ti/$^{48}$Ti = 0.0011 (0.068), $^{47}$Ti/$^{48}$Ti = 0.0013 (0.0025),
$^{49}$Ti/$^{48}$Ti = 0.0092 (0.082), and $^{50}$Ti/$^{48}$Ti =
0.31 (0.52).  Note the high relative abundance of $^{50}$Ti for W7 from
Nomoto et al. in contrast to the lower abundances for the same model
adopted by TWW95 and K06. One supposes that the $^{50}$Ti
yield is critically dependent on some adopted (and changing) nuclear
reaction rates in the reaction network. 

Hughes et al. (2008) predicted the evolution of the titanium isotopic
ratios for a dual-infall model. In this model, the halo forms by infall
of primordial gas with the disk formation and evolution accompanied
by a second infall episode with gas enriched by ejecta from halo
stars. The principal effect of this second episode on the predicted
evolution of the titanium isotopes is to increase the
$^{50}$Ti/$^{48}$Ti ratio in the disk. This increase is
presumably due to $^{50}$Ti production from Type Ia supernovae
because their model assumes a delay in the onset of the infall. The
resultant predicted isotopic ratios are quite similar to those
by Prantzos shown in Figures 7 and 8. (Hughes et al. adopt Nomoto
et al.'s (1997) yields for Type Ia supernovae.)

These GCE models neglect contributions from AGB stars, low mass
stars that do not die as supernovae but provide $s$-process and
other products. Operation of the neutron-capture $s$-process 
provides not only `heavy' nuclides (e.g., Sr and Ba) but also redistributes
the nuclides such as the Ti isotopes lighter than the Fe-peak. The Ti
isotopes most affected by the $s$-process in AGB stars appear to be
$^{49}$Ti and $^{50}$Ti; $^{50}$Ti is neutron magic ($N$=28) with a
small neutron-capture cross-section and, thus, is raised in 
abundance by the $s$-process. Calculations reported by Lugaro et al.
(1999) for a solar metallicity AGB star show that the production
factor of $^{50}$Ti is less than 10 for a production factor
of $s$-process `heavy' nuclides of 200-300. Thus, addition of AGB stars
to a GCE prescription to account for the abundances of $s$-process products
such a Sr and Ba is not expected to result in major alterations to the
Ti isotopic ratios predicted by TWW95, GP00, K06, and Hughes
et la. (2008). Modifications of Ti isotopic ratios attributed to
$s$-process operation in AGB stars are seen in presolar SiC grains
(Lugaro et al. 1999; Huss \& Smith 2007).

\subsection{Observed trends and predictions}

Observed ratios $^{i}$Ti/$^{48}$Ti and $^i$Ti/$^{46}$Ti
 are compared in
Figure 7 and 8, respectively, with the recent predictions from
Kobayashi and Prantzos. The predictions account rather well for
the observations. Indeed, the latter may be said to
fit three of the four ratios in Figure 7 well with the exception
being the $^{47}$Ti/$^{48}$Ti ratio that is underpredicted by
a factor of two or less. The former consistently underpredict the
abundances of the lesser abundant isotopes relative to $^{48}$Ti but,
except again for $^{47}$Ti/$^{48}$Ti, are within a factor of two of the
measured ratios.
In summary, the recent
predictions are an improvement on those by TWW95 mentioned
in introductions to the paper and the discussion.  These improvements
are
largely a reflection of changes to the adopted yields. The relatively
poor fit in Figure 7 to the observed 
 $^{47}$Ti/$^{48}$Ti ratios presumably arises from
a too low prediction for the yield of $^{47}$Ti from Type II supernovae,
primarily.  All in all, the degree of concordance in Figures 7 and 8 between
prediction and observation is a pleasing achievement.

Dissection of the reasons for the now (slight)
 disagreements between observed and predicted
isotopic ratios must ultimately account  for the larger failure of the
GCE models to predict the observed run of [Ti/Fe] with [Fe/H].
 The
predictions, as noted in the Introduction, match quite well the
shape of the observed run but predict values of [Ti/Fe] that are
about 0.4 dex less than observed at all [Fe/H].
A straight increase in yields of $^{48}$Ti but not other isotopes
quite obviously will completely destroy the agreement between
the predicted and observed isotopic ratios. 

 The simplest empirical
solution to this discrepancy for [Fe/H] $\leq -1$ 
 is to suppose that production of Ti by Type II supernovae
has been underestimated by about 0.4 dex. Constraints on this suggestion
could be provided were the isotopic ratios known for these
metal-poor stars (see below). For [Fe/H] $\geq -1$, both Type II
and Type Ia supernovae contribute to the [Ti/Fe] versus [Fe/H]
relation.  In this regime,
 enhanced Ti production must be achieved 
with $^{48}$Ti, the dominant isotope,  but no more than 
modest changes in the yields of the less abundant isotopes can be
tolerated if the measured isotope ratios in Figures 7 and 8
are to also fit. Is the present underproduction of $^{47}$Ti a
clue to how to make up the 0.4 dex discrepancy?  
In addition, theoretical proposals for
 achieving enhanced yields of Ti cannot seriously affect
presently predicted yields of 
Mg, Si, and Ca for which GCE models (e.g., GP00 and K06) reproduce well their
variation of [Element/Fe] versus [Fe/H]. The examples of Mg, Si, and Ca
also preclude the most naive way to reconcile the observed
and predicted runs of [Ti/Fe] with [Fe/H], i.e., invocation of  a 0.4 dex
reduction in the yield of Fe from Type II supernovae.
Theoreticians may ask - Is it at all possible that there's
a systematic error in the [Ti/Fe] estimates yet to be uncovered?
This seems unlikely given that very similar results are obtained from
samples of dwarfs and giants.

\section{Concluding Remarks}

Several potential observational tests of GCE and stellar nucleosynthesis
afforded by measurements of titanium isotopic ratios are left
unexplored by our initial foray.
Exploration calls for accurate measurement not only of the
titanium isotopic ratios but also of the metallicity. The key to
accurate metallicities is surely through observation and
analysis of infrared (J, H, and K band) spectra, as Martinache
et al. (2009) and Rojas-Ayala \& Lloyd (2009) are about to
demonstrate. 

Most notably, our observations do not extend to
the lower metallicities needed to isolate the contribution
from Type II supernovae alone, say [Fe/H] $< -1$. In particular,
the prediction that the low metallicity Type II supernovae are
inefficient producers of $^{50}$Ti has not been subjected to a clear
test. The isotope $^{50}$Ti is 
 present in GJ699 (Barnard's star) to which
we assign [Fe/H] $\simeq -0.8$ and seemingly too in LHS 178 with
[Fe/H] $\simeq -1$ for which predictions imply a fractional
abundance $^{50}$Ti/$^{48}$Ti of less than about one (Kobayashi)
 to three (Prantzos) per cent is
expected but $^{50}$Ti/$^{48}$Ti of six per cent is measured
for GJ699 and possibly a similar fraction for LHS178.

Exploration of the domain [Fe/H] $\leq -1$ is highly desirable.
The predicted  absence of $^{50}$Ti needs to be confirmed. 
The other three less abundant isotopes have predicted abundances of a
few per cent at [Fe/H]$=-2$, and $^{47}$Ti has even a 60\% increase
(relative to $^{48}$Ti) as [Fe/H] declines from 0 to $-2$ in Kobayashi's
calculations.
At [Fe/H]$=-2$, K06 predict $^{46}$Ti/$^{48}$Ti$\simeq0.046$,
$^{47}$Ti/$^{48}$Ti$\simeq 0.053$, and $^{49}$Ti/$^{48}$Ti$\simeq0.034$
with a negligible amount of $^{50}$Ti ($^{50}$Ti/$^{48}$Ti$<0.003$). 
Even at [Fe/H]$=-3$, these ratios, $^{50}$Ti excepted, are little
changed from the values at [Fe/H]$=-2$. Prantzos predicts lower
relative abundances but also a negligible amount of $^{50}$Ti:
$^{46}$Ti/$^{48}$Ti $\simeq$ 0.013, $^{47}$Ti/$^{48}$Ti $\simeq 0.005$,
$^{49}$Ti/$^{48}$Ti $\simeq 0.035$ for [Fe/H] $\leq -1.5$.

Exploration of  the range [Fe/H] $< -1$
will call for cooler stars than LHS 178 in order that lower
temperature may compensate for the weaker TiO bands. Few such
targets are yet known. Future discoveries of very metal-poor
cool dwarfs 
will be faint and access to high-resolution spectrographs on very large
telescopes will be desirable.
In fact, the optimum region for detection and measurement of the
$^{50}$Ti isotope is likely not the 0-0 band of the $\gamma$-system
that demands high spectral resolution
but  fortunately, as noted by Clegg et al. (1979),
 the 0-1 bandhead of the system
for which the $^{50}$TiO head falls to the blue of the
red-degraded R$_3$ bandhead by about 3 \AA\ with the $^{49}$TiO
head midway between the $^{50}$TiO and the $^{48}$TiO bandheads. A full
exploitation of this (and other potential indicators of the $^{50}$Ti
isotope) bandhead will require  $^{48}$TiO lines of similar
strength to the $^{50}$TiO bandhead. Clegg et al. suggested the
0-0 $\delta$-system but other possiblities can most likely
be found among the rich set of TiO electronic systems.  
Exploration of the spectra of cool subdwarfs will certainly be
necessary to find TiO features cleanly identifiable apart from
lines of other molecules, particularly those molecules like
CaH that strengthen relative to TiO as lower and lower
metallicities are encountered.

A project not presently lacking for target stars is the search
for isotopic abundance differences between thin and thick
stars.  This calls for  large samples of stars from both populations.
The difference in [Ti/Fe]  at a given [Fe/H] in the range of
[Fe/H] overlap for the two populations is about 0.15 dex from
analysis of F and G dwarfs (Reddy et al.
2006). 
Accurate spectroscopy might reveal a difference in the
fractional abundances for these populations.  
To achieve the necessary accuracy, it will be necessary to
observe  $^{48}$TiO features of a comparable strength to the 
isotopic lines from the $\gamma$-system's 0-0 band,  as suggested above.
 Reddy et al. suggest that the difference in [Ti/Fe] between the
two populations may be largely
 attributable to the formation of the thin but not thick disk stars from gas
contaminated 
with Type Ia supernovae ejecta.  
If $^{50}$Ti is made in copious amounts (relative to $^{48}$Ti) as some
predictions of Type Ia supernovae suggest, the proposed contamination
seems difficult to reconcile with the
observation here that the thick disk stars have as much, if not more
$^{50}$Ti, as the thin disk stars.
However, interpretation is complicated by the prediction that the $^{50}$Ti
yield from Type II supernovae increases steeply as [Fe/H] increases from
$-1$ to higher values. This is just the [Fe/H] interval over which Type Ia
supernovae make an increasing contribution to composition of the
interstellar gas.

\bibliographystyle{natbib} 
%\bibliography{tiisotope.bbl}  

We thank Chiaki Kobayashi  for providing unpublished
predictions for the Galactic chemical evolution of Ti isotopes and
Luca Casagrande for sending us photometric estimates of the
stellar metallicities. We are especially grateful to Nikos
Prantzos for a helpful commentary on the GCE of titanium isotopes
and for providing unpublished calculations.
 We thank Carlos Allende Prieto, Thomas G. Barnes, III,
Fritz Benedict, and Chris Sneden for helpful discussions. 
JMC thanks Jacob Bean and Ian Roederer
for assistance and encouragement. 
We thank the referee for a helpful report.
This research has been
supported in part by a grant to DLL from the Robert A. Welch
Foundation of Houston, Texas.

%\end{document}

%\section{} 
%{\it Facilities:} \facility{Subaru}, \facility{HST (STIS)}}.

%% the file, uncomment the \includegraphics, \plottwo, and
%% If you need a layout that cannot be achieved with \includegraphics or
%% \plottwo, you can invoke the graphicx package directly with the
%% \includegraphics command or use \plotfiddle. For more information,
%% please see the tutorial on "Using Electronic Art with AASTeX" in the
%% documentation section at the AASTeX Web site,
%% http://www.journals.uchicago.edu/AAS/AASTeX.
%%
%% The examples below also include sample markup for submission of
%% supplemental electronic materials. As always, be sure to check
%% the instructions to authors for the journal you are submitting to
%% for specific submissions guidelines as they vary from

%% This example uses \includegraphics to include an EPS file scaled to
%% 80% of its natural size with \epsscale. Its caption
%% has been written to indicate that additional figure parts will be
%% available in the electronic journal.

\clearpage

%\begin{figure}
%%\plottwo{rv1327.ps}{hr1327.ps}
%\caption{Predicted isotopic ratios $^{i}$Ti/$^{48}$Ti versus [Fe/H] from
%GCE models by the unbroken line from
%Goswami \& Prantzos (2000, private communication) and by the dashed line from
%Kobayashi et al. (2006, private communication). Solar system isotopic
%abundances are indicated. 
%}
%\end{figure}

\begin{figure}
\includegraphics{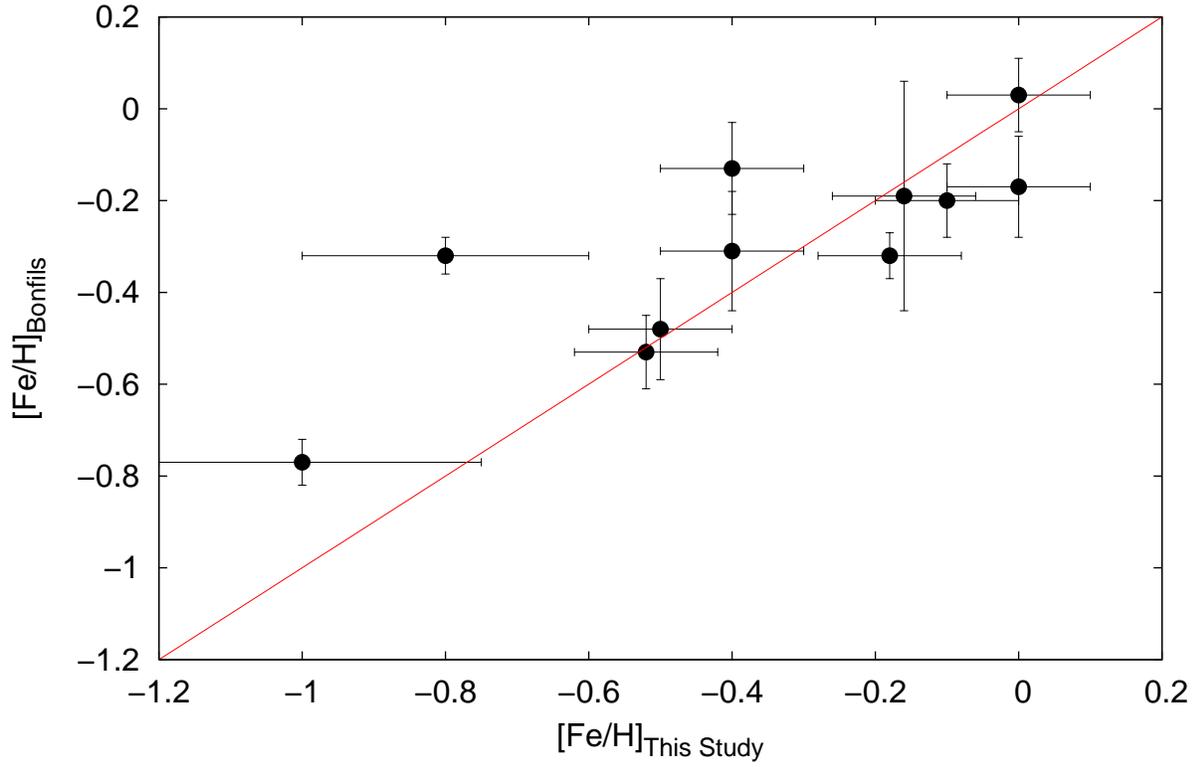}
\caption{A comparison between our spectroscopic [Fe/H]  and metallicities
derived from Bonfils et al.'s (2005) photometric calibration (their
equation 1). The red line indicates exact equality between the
results from the two methods.
}
\end{figure}

\begin{figure}
\includegraphics{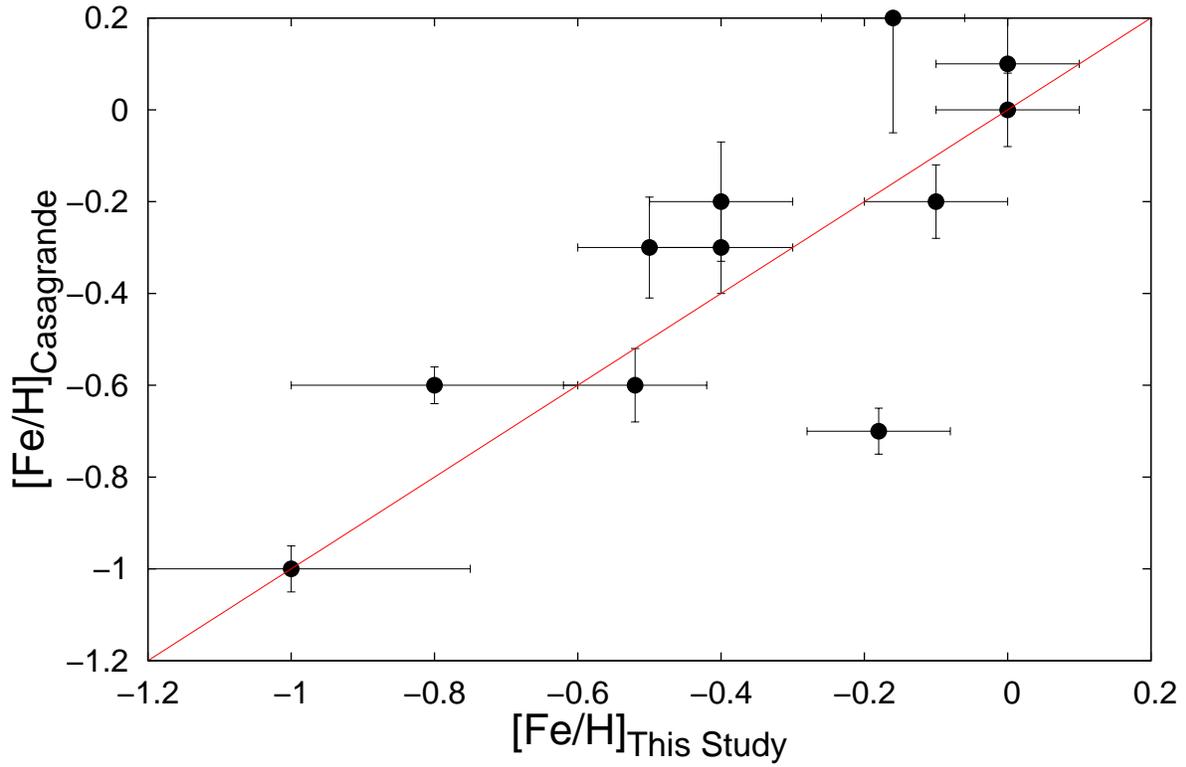}
\caption{A comparison between our spectroscopic [Fe/H]  and metallicities
derived from Casagrande et al.'s  (2008) photometric calibration
(Luca Casagrande, private communication).
 The red line indicates exact equality between the
results from the two methods.
}
\end{figure}

\begin{figure}
\includegraphics{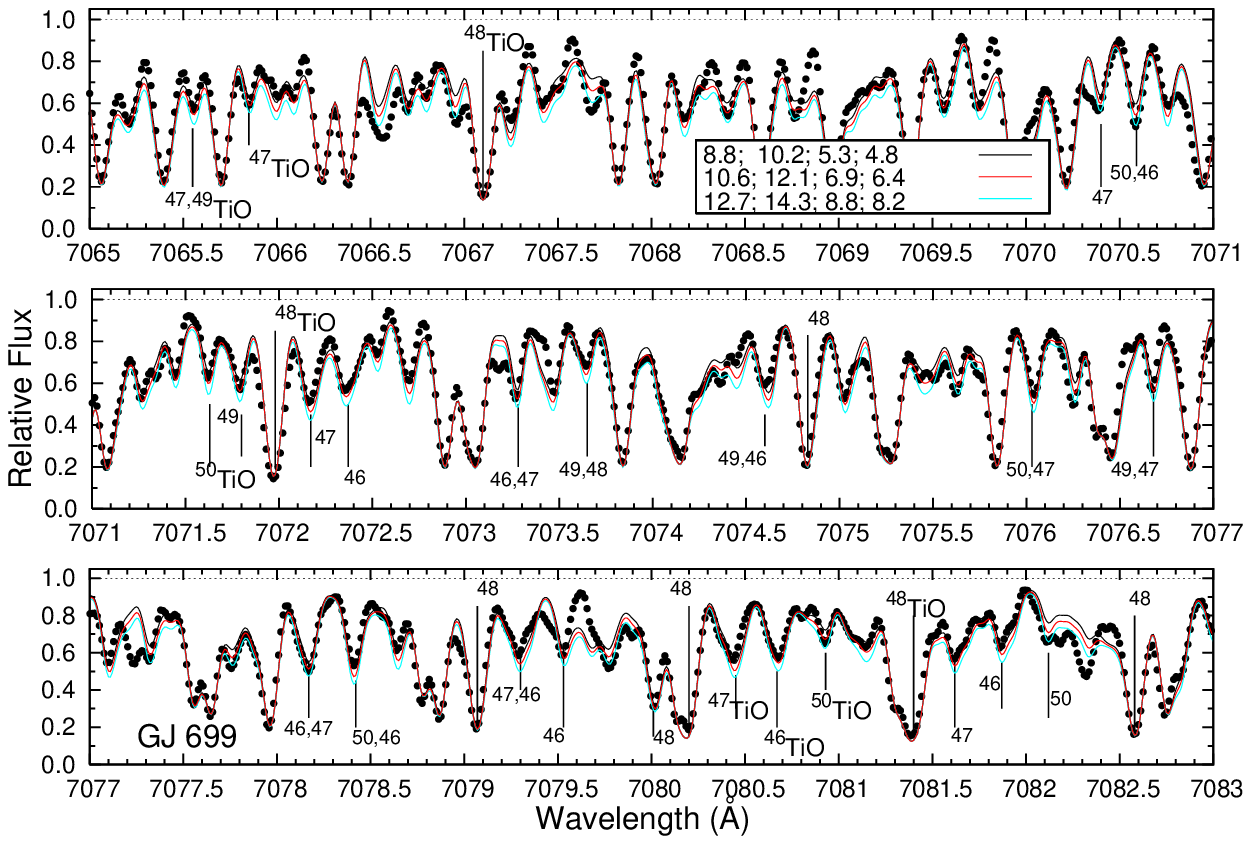}
\caption{The spectrum of GJ699 for 7065 \AA\ to 7083 \AA\ showing
absorption lines for all five varieties of $^i$TiO from the 0-0 band of the
$\gamma$-system. The strongest lines are from $^{48}$TiO. Weaker lines 
include lines from the other Ti isotopes with several key lines and blends
labelled by the mass number. Synthetic spectra for three
isotopic mixes are  shown with the key in the upper panel.
All three synthetic spectra fit the $^{48}$TiO lines.
The key to the isotopic mixes is given on the figure with left to
right the abundances of $^{46}$Ti, $^{47}$Ti, $^{49}$Ti, and
$^{50}$Ti expressed in per cent relative to the $^{48}$Ti abundance.
}
\end{figure}

\begin{figure}
\includegraphics{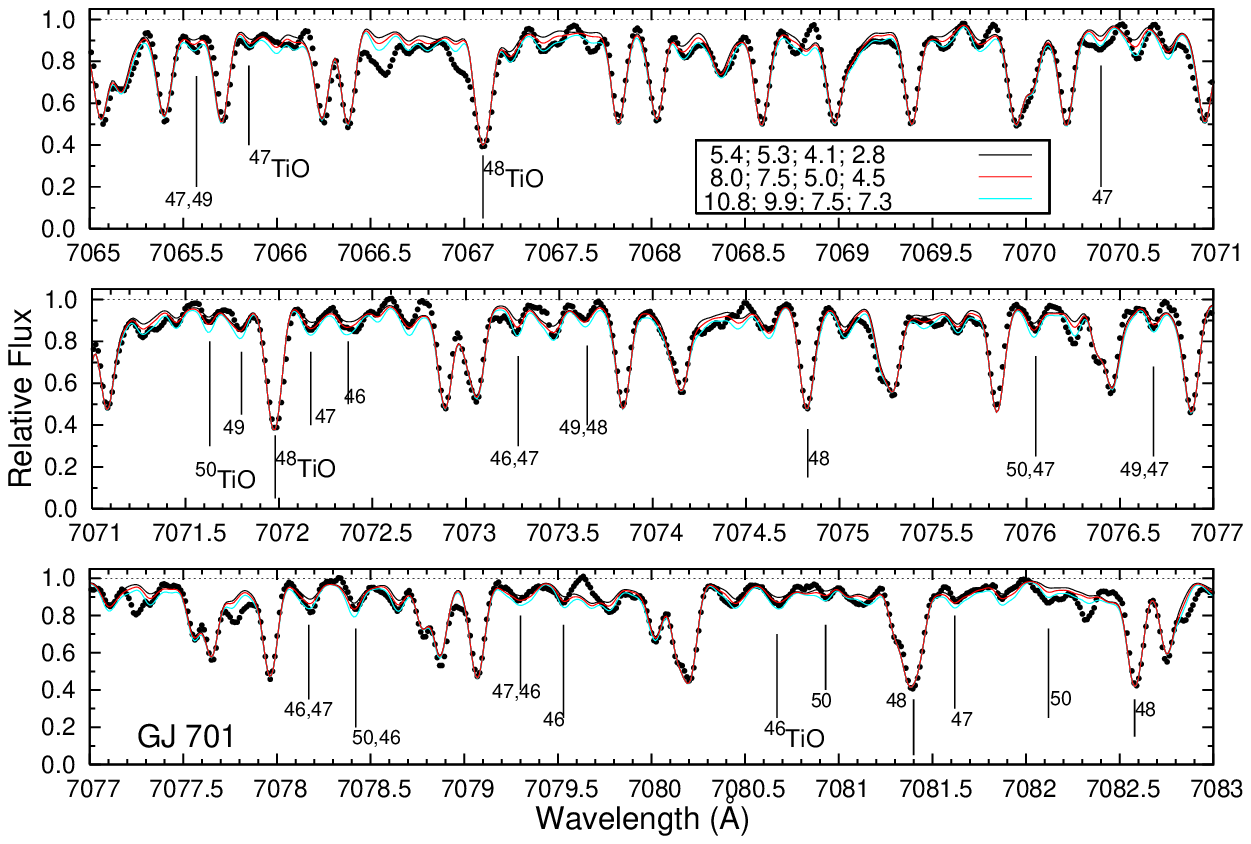}
\caption{The spectrum of GJ701 for 7065 \AA\ to 7083 \AA\ showing
absorption lines for all five varieties of $^i$TiO from the 0-0 band of the
$\gamma$-system. The strongest lines are from $^{48}$TiO. Weaker lines 
are from the other Ti isotopes with several key lines and blends
labelled by the mass number. Synthetic spectra for three
isotopic mixes are  shown with the key in the upper panel.
All three synthetic spectra fit the $^{48}$TiO lines.
The key to the isotopic mixes is given on the figure with left to
right the abundances of $^{46}$Ti, $^{47}$Ti, $^{49}$Ti, and
$^{50}$Ti expressed in per cent relative to the $^{48}$Ti abundance.
}
\end{figure}
 
\begin{figure}
\includegraphics{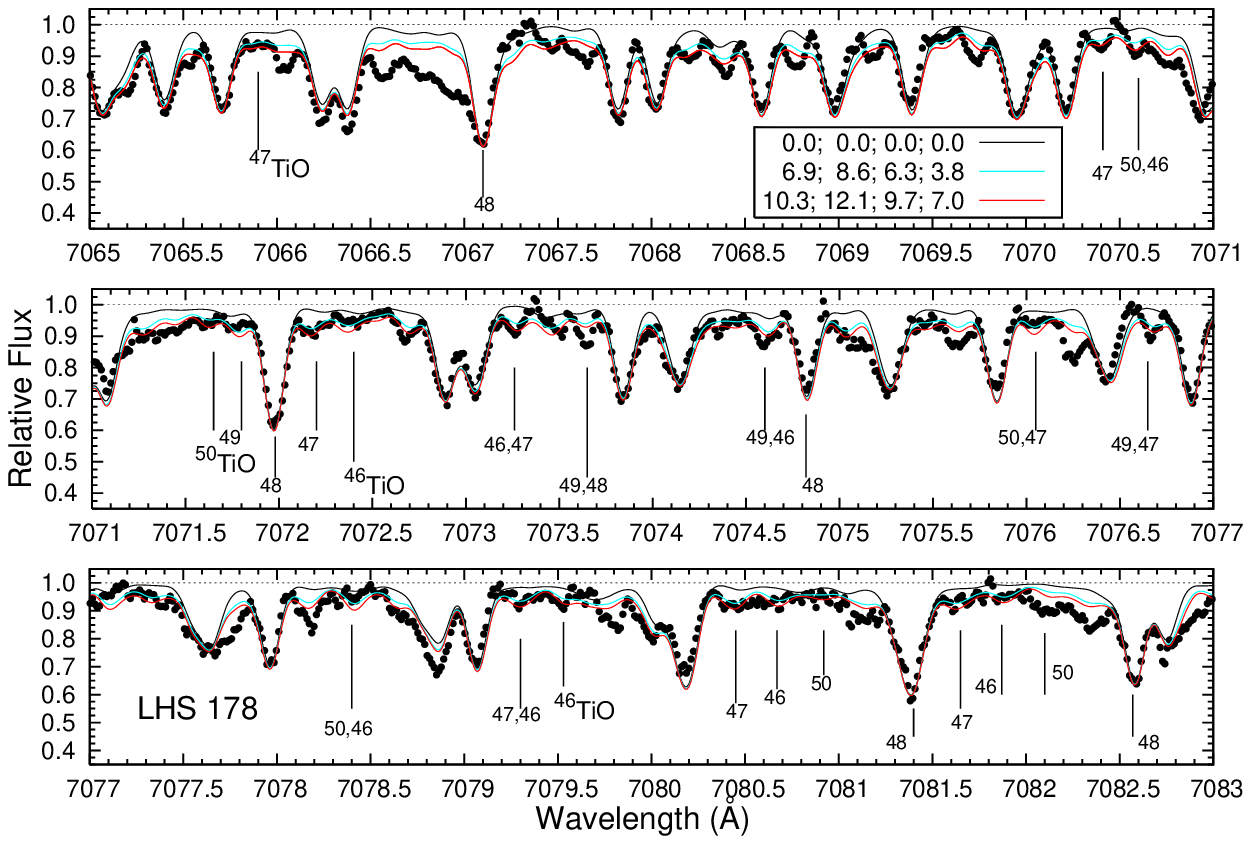}
\caption{The spectrum of LHS178 for 7065 \AA\ to 7083 \AA\ showing
absorption lines for all five varieties of $^i$TiO from the 0-0 band of the
$\gamma$-system. The strongest lines are from $^{48}$TiO. Weaker lines 
are from the other Ti isotopes with several key lines and blends
labelled by the mass number. Synthetic spectra for three
isotopic mixes are shown with the key in the upper panel.
All three synthetic spectra fit the $^{48}$TiO lines.
The key to the isotopic mixes is given on the figure with left to
right the abundances of $^{46}$Ti, $^{47}$Ti, $^{49}$Ti, and
$^{50}$Ti expressed in per cent relative to the $^{48}$Ti abundance.
The synthetic spectrum for pure $^{48}$TiO shows that the collective
contribution from the four lesser abundant Ti isotopes is present here.
}
\end{figure}
 
\begin{figure}
\includegraphics{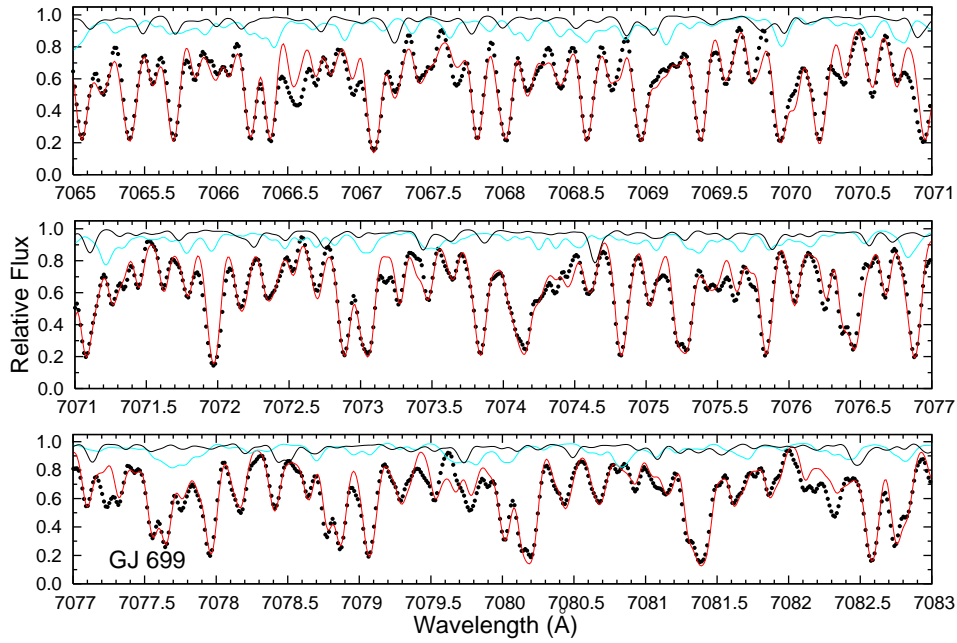}
\caption{The observed spectrum of GJ699 for 7075 \AA\ to 7083 \AA\ with 
synthetic spectra for three different selections of TiO lines: (i)
lines (all Ti isotopes) from the P, Q, and R branches of the
0-0 band of the TiO  $\gamma$-system (red line), (ii) satellite lines of the
0-0 band (light blue line), and (iii) lines of the $\Delta v = +1$ bands of the
$\gamma$-system (black line). 
}
\end{figure}

\begin{figure}
\includegraphics{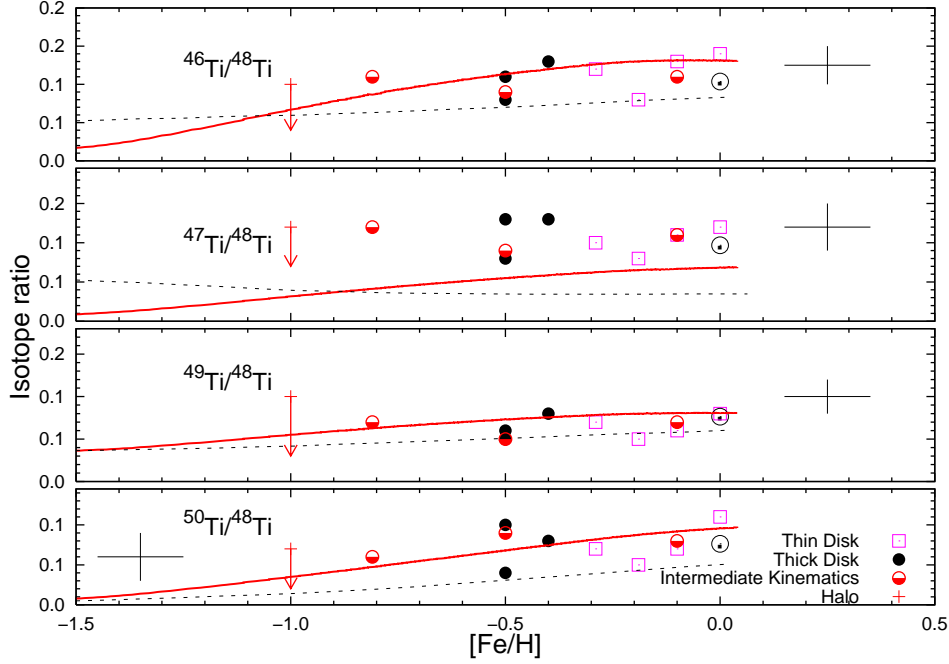}
\caption{Observed and predicted isotopic ratios $^i$Ti/$^{48}$Ti
over the [Fe/H] range 0 to $-1.5$. Stellar populations of our stars
are represented as follows: thick disk=filled circles, thin disk=unfilled
squares, halo=cross (arrow indicates an upper limit), and thick-thin disk = intermediate kinematics = half-filled circle.
 Predictions
are from Prantzos (private communication, red line) and 
Kobayashi (private communication,  black dashed line).
An indication of  representative uncertainty is given  by the cross in each panel.
Solar system ratios are denoted at [Fe/H]=0 by the standard symbol for the Sun. }
\end{figure}

\begin{figure}
\includegraphics{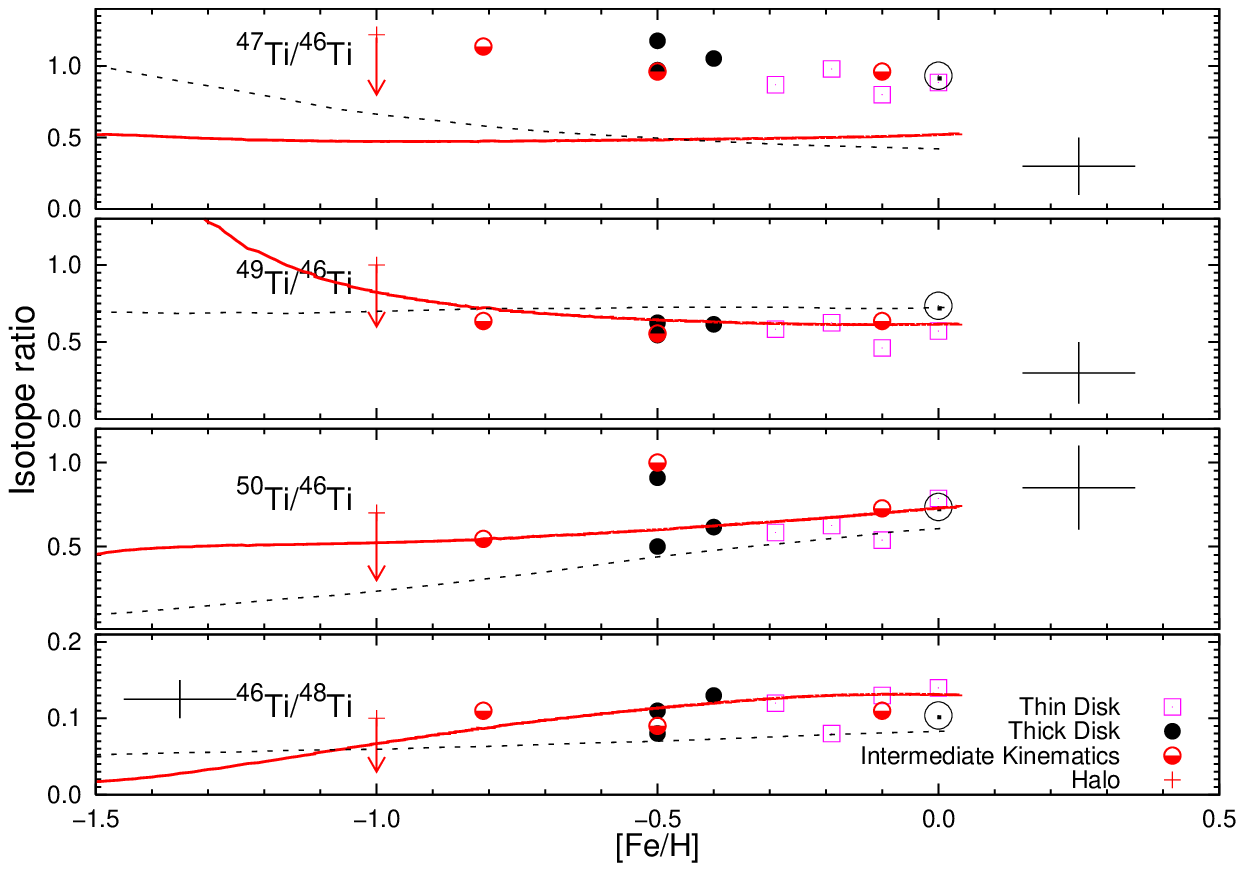}
\caption{Observed and predicted isotopic ratios $^{47}$Ti/$^{46}$Ti,
$^{49}$Ti/$^{46}$Ti, $^{50}$Ti/$^{46}$Ti, and $^{46}$Ti/$^{48}$Ti. 
Stellar populations of our stars
are represented as follows: thick disk=filled circles, thin disk=unfilled
squares, halo=cross (arrow indicates an upper limit), and thick-thin disk= intermediate kinematics=half-filled circle.
 Predictions
are from Prantzos (private communication, red line) 
and Kobayashi (private communication, (black dashed line). 
An indication of  representative uncertainty is given by the cross in each panel.
Solar system ratios are denoted at [Fe/H]=0 by the standard symbol for the Sun. }
\end{figure}

\clearpage

%TABLE1

  \pagestyle{empty}

  \setcounter{table}{0}
  \begin{table}
  \caption{The Observed Stars } \vspace{.25cm}
  \scriptsize
  \begin{tabular}{lrrll }
  \hline
Star & V &   Sp.Type & Population\\
\hline
GJ184 &  9.2 &  M0.5V & Thick disk \\
GJ215 & 8.3 &   K7V & Thin disk \\
GJ378 & 9.3 & M1V & Thick disk \\
GJ699 &  9.5 & M4V & Thin (65\%) - Thick (33\%) disk\\
GJ701 & 9.9 & M1V & Thin disk \\
GJ725A &  8.9 & M3V & Thin disk \\
GJ880 &  9.6 & M1.5V & Thin disk\\
GJ908 & 10.2 & M1V & Thin (24\%) - Thick (72\%) disk \\
LHS178 & 10.7 & M1V & Halo \\
LHS1226 & 9.5 & M0.5V & Thin (32\%) - Thick (66\%) disk  \\
LHS2018 & 7.9 & K7V & Thick disk\\
\hline
    \end{tabular} \\

   \end{table}
%\end{document}

\clearpage
%TABLE2

% \documentstyle[apjpt4]{article}
% \begin{document}

  \pagestyle{empty}

  \setcounter{table}{1}
  \begin{table}
  \caption{Stellar Parameters } \vspace{.25cm}
  \scriptsize
  \begin{tabular}{lccccc}
  \hline
Star & T$_{\rm eff}$ & log g & [Fe/H] &$ \xi$ & $\zeta$\\
      & (K) & (cgs) & & (km s$^{-1}$) & (km s$^{-1}$)\\
\hline
GJ184 & 3700 & 4.7 & -0.5 & 1.0 & 2.0  \\
GJ215 &  3900 & 4.5 & -0.1 & 1.0 & 1.0 \\
GJ378 &  3600 & 4.6 & -0.4 & 1.4 & 0.7\\
GJ699  &  3134 & 5.1 & -0.8 & 0.6 & 1.4 \\
GJ701 &  3680 & 4.8 & -0.2 & 1.0 & 0.8\\
GJ725A &  3400 & 4.9 & -0.3 & 1.1 & 1.0\\
GJ880 &  3640 & 4.7 &\  0.0 & 1.0 & 1.5 \\
GJ908 & 3550 & 4.8 & -0.5 & 0.8 & 0.8 \\
LHS178 & 3600 & 5.0 & -1.0 & 1.0 & 1.0  \\
LHS1226 & 3900 & 4.8 & -0.1 & 0.6 & 1.4 \\
LHS2018 &  3750 & 4.7 & -0.5 & 1.0 & 1.0 \\
\hline
    \end{tabular} \\
     \end{table}
% \end{document}

\clearpage

%TABLE3

  \clearpage
\setcounter{table}{2}
\begin{singlespace}
\begin{deluxetable}{lrrrrrrrr}
\tablecolumns{8}
\tablewidth{0pc}
\tablecaption{Metallicity study}
\tablehead{\\[-7mm]
\colhead{}& \multicolumn{6}{c}{[Ti/H]}& \colhead{\underline{$\rm 
[Fe/H]$}}\\[-0.5mm]
\cline{2-7} \\[-3mm]
\colhead{Star}& \multicolumn{5}{c}{Ti \textsc{i}}& 
\colhead{\underline{TiO}}& \colhead{}\\[-.8mm]
\cline{2-6} \\[-2mm]
\colhead{}& \colhead{8438.9\AA}& \colhead{8450.9\AA}& \colhead{8457.1\AA}& 
\colhead{8467.1\AA}& \colhead{Mean}& \colhead{}& \colhead{}
}
\startdata
GJ 184&   -0.21& -0.40& -0.45& -0.28& -0.34& -0.30 & -0.5\\
GJ 215& 0.0& -0.05& -0.25& 0.05& -0.06& -0.13& -0.1\\
GJ 378& -0.09& -0.19& -0.34& -0.11& -0.18& -0.28&  -0.4\\
GJ 699& -0.68& -0.64& -0.61& -0.49& -0.61& -0.63& -0.8\\
GJ 701&  -0.20& -0.25& -0.25& -0.10& -0.20& -0.18 & -0.2\\
GJ 725A&  -0.18& -0.23& -0.21& -0.04& -0.17& -0.30 & -0.3    \\*
%GJ 834 A & T=3800& Thin Disk&$\xi$=1.0 &$\zeta$=1.0 & \\
%$ \log \epsilon_{Ti}$ &  5.4& 5.3& 5.36& 5.21& 5.22\\
%$\rm [Ti/H]$ && \\
GJ 880& 0.10& 0.02& -0.07& 0.10 & 0.04&  -0.13&   0.0\\
GJ 908&  -0.40& -0.44& -0.55& -0.35& -0.44& -0.38&  -0.5\\*
LHS 178 & -1.00& -0.95& -0.95& -0.75& -0.97& -0.55& -1.0 \\
LHS 1226 & -0.12& -0.22& -0.20& -0.06& -0.10& -0.05& -0.1 \\
LHS 2018& -0.36& -0.46& -0.50& -0.35& -0.37& -0.39& -0.5 \\
\enddata
\label{metallicity}
\end{deluxetable}
\end{singlespace}

\clearpage
%TABLE4

%\documentstyle[apjpt4]{article}
% \begin{document}

  \pagestyle{empty}

  \setcounter{table}{3}
  \begin{table}
  \caption{Comparisons with Bonfils et al. (2005) and Casagrande et al.
(2008) } \vspace{.25cm}
  \scriptsize
  \begin{tabular}{lcccccc}
  \hline
Star &\multicolumn{2}{c}{T$_{\rm eff}$} & & \multicolumn{3}{c}{[Fe/H]}  \\
\cline{2-3}\cline{5-7}
  & CL$^a$ & CFB$^b$ & & B$^c$  & CFB$^b$ & CL$^a$\\

\hline
GJ184 & 3700 &   3690 && -0.3 & -0.2 & -0.5 \\
GJ215 &  3900 & 3950 & &-0.2 & 0.1 & -0.1 \\
GJ378 &  3600 & 3590 && -0.1 & -0.3 & -0.4 \\
GJ699  &  3134 & 3150 && -0.3 & -0.6 & -0.8 \\
GJ701 &  3680 & 3560 && -0.2 & -0.2 & -0.2 \\
GJ725A &  3400 & 3300 && -0.3 & -0.7 & -0.3 \\
GJ880 &  3640 & 3540 && 0.0 & 0.0 & 0.0 \\
GJ908 & 3550 & 3560 && -0.5 & -0.6 & -0.5 \\
LHS178 & 3600 & 3500 && -0.8 & -1.0 & -1.0  \\
LHS1226 & 3900 &  3700 && -0.2 & -0.3 & -0.1 \\
LHS2018 &  3750 &  3960 && -0.5 & 0.2 & -0.5 \\
\hline
    \end{tabular} \\
    $^a$ This paper\\
    $^b$ Casagrande et al. (2008)\\
    $^c$ Bonfils et al. (2005)\\
     \end{table}
% \end{document}

\clearpage
%TABLE5

%\documentstyle[apjpt4]{article}
%\begin{document}

  \pagestyle{empty}

  \setcounter{table}{4}
  \begin{table}
  \caption{Primary TiO Lines} \vspace{.25cm}
  \scriptsize
  \begin{tabular}{lllcc }
  \hline
  $\lambda$(\AA)& Transition & Isotopes  \\
   \hline
7058.7 & R$_{3}$(33) & 48 \\
7060.4 & Q$_{3}(10)$ R$_{3}(36)$ & 48\\
7062.5 & Q$_{3}(15)$ R$_{3}(39)$ & 47\\
7065.9 & Q$_3$(21) &   47\\
7067.1 & P$_3(12)$ Q$_3(23)$ R$_3(45)$ & 48\\
7070.4 & Q$_3(27)$ & 47\\
7070.6 & Q$_3(27)$ & 46, 50 \\
7071.6 & Q$_3(29)$ P$_3(17)$ R$_3(50)$ &  50\\
7071.8 & Q$_3(29)$ P$_3(17)$ R$_3(50)$ & 49\\
7072.0 & Q$_3(29)$ P$_3(17)$ R$_3(50)$ &  48\\
7072.4 & Q$_3(29)$ P$_3(17)$ R$_3(50)$ &  46\\
7072.2 & Q$_3(29)$ P$_3(17)$ R$_3(50)$ &  47\\
7073.3 & Q$_3(30)$ P$_3(18)$ R$_3(51)$ &  46, 47\\
7073.7 & Q$_3(31)$ &  49\\
7074.9 & Q$_3(32)$  & 48\\
7076.1 & Q$_3(33)$ P$_3(21)$ R$_3(54)$ &  47, 50\\
7076.7 & Q$_3(34)$ R$_3(54)$ &  47, 49\\
7078.2 & Q$_3(35)$ R$_3(55)$ &  46, 47\\
7078.4 & Q$_3(35)$ P$_3(23)$ R$_3(56)$ & 46, 50\\
7079.3 & Q$_3(36)$ P$_3(23)$ R$_3(56)$ &  46, 47\\
7079.5 & Q$_3(36)$ &  46\\
7080.5 & Q$_3(37)$ R$_3(57)$ &  47\\
7080.7 & Q$_3(37)$ R$_3(57)$ &  46\\
7080.9 & Q$_3(38)$ P$_3(25)$ R$_3(58)$ &  50\\
7081.4 & Q$_3(38)$ P$_3(25)$ R$_3(58)$ &  48\\
7081.7 & Q$_3(38)$ P$_3(25)$ R$_3(58)$ &  47\\
7081.9 & Q$_3(38)$ P$_3(25)$ R$_3(58)$ &  46\\
7082.6 & Q$_3(39)$ P$_3(26)$           &  48\\
7083.2 & Q$_3(39)$ P$_3(26)$ R$_3(59)$ &  46, 47\\
7083.4 & Q$_3(40)$                     &  50\\

\hline
    \end{tabular} \\

     \end{table}
%\end{document}

\clearpage
%TABLE6

%\documentstyle[apjpt4]{article}
%\begin{document}

  \pagestyle{empty}

  \setcounter{table}{5}
  \begin{table}
  \caption{$^i$TiO Abundances}\vspace{.25cm}
  \scriptsize
  \begin{tabular}{llrrrrrr}
  \hline
  &&& &Star &&\\
\cline{3-7}
$\lambda$(\AA)  & Isotope & GJ699 & & GJ701 & & LHS178\\
\cline{3-3}\cline{5-5}\cline{7-7}
& & log $\epsilon(^{i}Ti)  $ & & log $\epsilon(^{i}Ti)$ & &  log
$\epsilon(^{i}Ti)   $\\
\hline
7072.4 & 46 & 3.18 && 3.64 & & 3.11 \\
7079.5 & 46 & 3.21 && 3.70 && \nodata \\
7080.7 & 46 & 3.18 && 3.70 && 3.11 \\
7081.9 & 46 & 3.17 && \nodata  && 3.21\\
7073.3 & 46[47]$^{a}$ & 3.29 && 3.76 && \nodata \\
7078.2 & 46[47] & 3.26 && 3.76 && \nodata \\
7079.3 & 46[47] & 3.18 && 3.46 && 3.11\\
7083.2 & 46[47] & 3.23 && 3.64 && \nodata\\
&&&&\\
& {\bf Mean ($^{46} $Ti  )} &  3.20$\pm$0.1& & 3.62 $\pm$ 0.06 & & 
3.35$\pm$0.09\\
&&&&\\
7062.5 & 47 & 3.21 && 3.46 && 3.35\\
7065.9 & 47 & 3.31 && 3.70 && 3.35\\
7070.4 & 47 & 3.35 && 3.85 && 3.51\\
7072.2 & 47 & 3.21 && 3.64 && 3.44\\
7080.5 & 47 & 3.21 && \nodata && 3.44\\
7081.7 & 47 & 3.21 && 3.56 && 3.44\\
&&&&\\
&  {\bf Mean ($^{47}$Ti )} & 3.25 $\pm 0.1$ && 3.77$ \pm 0.07$  & &
3.43$\pm 0.07$ \\
&&&\\
7065.6 & 49[47] & 2.99 && 3.56 && \nodata\\
7071.8 & 49    &  3.03 && 3.46 && 3.15\\
7073.7 & 49 & 2.99 && 3.34 && \nodata \\
7074.6 & 49[46] & \nodata && \nodata && 3.32\\
7076.7 & 49[47] & 3.03 && 3.56 && 3.40\\
&&&\\
& {\bf Mean ($^{49}$Ti )}  & 3.01$\pm 0.08$ && 3.44 $\pm 0.11$   & & 
3.34$\pm
0.1$\\
  &&&\\
  7071.6 & 50 & 2.95 && 3.37 && 3.10 \\
  7080.9 & 50 & 3.04 && 3.53 && 3.20\\
  7083.4 & 50 & 2.90 && 3.43 && 3.20\\
  7076.1 & 50[47] & 2.84 && 3.43 && \nodata \\
  7078.4 & 50[46] & 2.69 && 3.43 && \nodata \\
  7070.6 & 50[46] & 3.23 && \nodata && 3.29\\
  &&&&\\
&  {\bf Mean ($^{50}$Ti )} & 2.98$\pm 0.1$ && 3.42 $\pm 0.11$  & & 
3.19$\pm$0.15\\
&&&\\
7058.7 & 48 & 4.23 && 4.74 && 4.38\\
7060.4 & 48 & 4.18 && 4.74 && 4.41\\
7067.1 & 48 & 4.12 && 4.74 && 4.34\\
7072.0 & 48 & 4.23 && 4.74 && 4.31\\
7074.9 & 48 & 4.18 && 4.74 && 4.27\\
7081.4 & 48 & 4.08 && 4.74 && 4.38\\
7082.6 & 48 & 4.14 && 4.74 && 4.35\\
&&&\\
& {\bf Mean ($^{48}$Ti)} & 4.17$\pm 0.16$ && 4.74 $\pm 0.03$ & & 4.35$\pm 
0.1$\\
&&&\\
\hline\\
  \end{tabular} \\
$^a$  The isotope in square brackets is given its mean abundance in
estimating the abundance of the other isotope.
  \end{table}

\clearpage
%TABLE7

\pagestyle{empty}

\setcounter{table}{6}
\begin{table}
\caption{Isotopic Ti Abundances}\vspace{.25cm}
\scriptsize
\begin{tabular}{lcccccc}
\hline
Star & \multicolumn{6}{c}{log $\epsilon\ (^i $Ti)}\\
\cline{2-7}
& i=46 & 47 & 48 & 49 & 50 & Total \\
\hline
GJ184 & 3.43 $\pm$ 0.05 & 3.42 $\pm$ 0.05 & 4.53 $\pm$ 0.05 & 3.20 $\pm$ 
0.10 & 3.18 $\pm$ 0.16 & 4.63\\
GJ215 & 3.77 $\pm$ 0.03 & 3.68 $\pm$ 0.04 & 4.65 $\pm$ 0.03 & 3.44 $\pm$ 
0.08 & 3.50 $\pm$ 0.05 & 4.79\\
GJ378 & 3.63 $\pm$ 0.04 & 3.65 $\pm$ 0.04 & 4.53 $\pm$ 0.06 & 3.41 $\pm$ 
0.05 & 3.44 $\pm$ 0.04 & 4.68\\
GJ699 & 3.20 $\pm$ 0.10 & 3.25 $\pm$ 0.10 & 4.17 $\pm$ 0.16 & 3.01 $\pm$ 
0.10 & 2.98 $\pm$ 0.10 & 4.30\\
GJ701 & 3.62 $\pm$ 0.06 & 3.61 $\pm$ 0.07 & 4.74 $\pm$ 0.03 & 3.44 $\pm$ 
0.11 & 3.42 $\pm$ 0.11 & 4.84\\
GJ725A & 3.59  $\pm$ 0.07 & 3.53 $\pm$ 0.07 & 4.53 $\pm$ 0.12 & 3.37 $\pm$ 
0.03 & 3.35 $\pm$ 0.03 & 4.66\\
GJ880 & 3.53 $\pm$ 0.02 & 3.48 $\pm$ 0.03 & 4.40 $\pm$ 0.03 & 3.30 $\pm$ 
0.04 & 3.43 $\pm$ 0.03 & 4.56\\
GJ908 & 3.38 $\pm$ 0.06 & 3.37 $\pm$ 0.07 & 4.42 $\pm$ 0.10 & 3.14 $\pm$ 
0.09 & 3.35 $\pm$ 0.07 & 4.54\\
LHS178 & 3.35 $\pm$ 0.09 & 3.43 $\pm$ 0.07 & 4.35 $\pm$ 0.06 & 3.34 $\pm$ 
0.10 & 3.19 $\pm$ 0.15 & 4.49\\
LHS1226 & 3.75 $\pm$ 0.05 & 3.71 $\pm$ 0.05 & 4.69 $\pm$ 0.08 & 3.52 $\pm$ 
0.06 & 3.62 $\pm$ 0.06 & 4.83\\
LHS 2018 & 3.40 $\pm$ 0.04 & 3.62 $\pm$ 0.03 & 4.51 $\pm$ 0.03 & 3.32 
$\pm$ 0.04 & 3.50 $\pm$ 0.04 & 4.65\\
\hline
\end{tabular} \\
\end{table}

\clearpage
%TABLE8

%\begin{document}

  \pagestyle{empty}

  \setcounter{table}{7}
  \begin{table}
  \caption{Isotopic Fractions f($i$) = $^i$Ti/$\Sigma ^i$Ti } 
\vspace{.25cm}
  \scriptsize
  \begin{tabular}{lcccccc}
  \hline
Star & [Fe/H] & f(46) & f(47) & f(48) & f(49) & f(50)\\
\hline
GJ184 &   -0.5 & 0.064 $\pm$ 0.009 & 0.062 $\pm$ 0.010 & 0.801 $\pm$ 0.098 
& 0.037 $\pm$ 0.016 & 0.036 $\pm$ 0.017\\
GJ215 &   -0.1 & 0.096 $\pm$ 0.008 & 0.078 $\pm$ 0.011 & 0.729 $\pm$ 0.060 
& 0.045 $\pm$ 0.013 & 0.052 $\pm$ 0.015\\
GJ378 &   -0.4 & 0.089 $\pm$ 0.009 & 0.093 $\pm$ 0.011 & 0.707 $\pm$ 0.101 
& 0.054 $\pm$ 0.011 & 0.057 $\pm$ 0.015 \\
GJ699 &   -0.8 & 0.079 $\pm$ 0.018 & 0.088 $\pm$ 0.020 & 0.735 $\pm$ 0.275 
& 0.051 $\pm$ 0.015 & 0.047 $\pm$ 0.018\\
GJ701 &   -0.2 & 0.061 $\pm$ 0.006 & 0.059 $\pm$ 0.008 & 0.801 $\pm$ 0.059 
& 0.040 $\pm$ 0.010 & 0.038 $\pm$ 0.015\\
GJ725A &-0.3 & 0.085 $\pm$ 0.014 & 0.074 $\pm$ 0.014 & 0.741 $\pm$ 0.199 & 
0.051 $\pm$ 0.013 & 0.049 $\pm$ 0.017\\
GJ880 &    0.0 & 0.094 $\pm$ 0.007 & 0.083 $\pm$ 0.011 & 0.694 $\pm$ 0.054 
& 0.055 $\pm$ 0.010 & 0.074 $\pm$ 0.015\\
GJ908 &   -0.5 & 0.069 $\pm$ 0.010 & 0.068 $\pm$ 0.012 & 0.759 $\pm$ 0.183 
& 0.040 $\pm$ 0.013 & 0.065 $\pm$ 0.017\\
LHS178 & -1.0 & 0.072 $\pm$ 0.014 & 0.087 $\pm$ 0.016 & 0.721 $\pm$ 0.107 
& 0.070 $\pm$ 0.018 & 0.050 $\pm$ 0.021\\
LHS1226& -0.1 & 0.084 $\pm$ 0.010 & 0.076 $\pm$ 0.011 & 0.729 $\pm$ 0.142 
& 0.049 $\pm$ 0.011 & 0.062 $\pm$ 0.016\\
LHS2018& -0.5 & 0.057 $\pm$ 0.009 & 0.094 $\pm$ 0.010 & 0.731 $\pm$ 0.053 
& 0.047 $\pm$ 0.010 & 0.071 $\pm$ 0.015\\
\hline
    \end{tabular} \\
   \end{table}

\end{document}